\documentclass[fleqn,usenatbib]{mnras}

\usepackage[T1]{fontenc}

\DeclareRobustCommand{\VAN}[3]{#2}
\let\VANthebibliography\thebibliography
\def\thebibliography{\DeclareRobustCommand{\VAN}[3]{##3}\VANthebibliography}

\usepackage{graphicx}	
\usepackage{amsmath}	
\usepackage{amssymb}	
\usepackage{pdflscape}

\usepackage{xcolor}

\title[ETV analysis of {\em OGLE-IV} EBs II.]{Eclipse timing variation analysis of {\em OGLE-IV} eclipsing binaries toward the Galactic Bulge.\\ II.
Short periodic triple stellar systems}

\author[T. Hajdu et al.]{
T. Hajdu$^{1,2,3}$\thanks{E-mail: hajdu.tamas@csfk.org},
T. Borkovits$^{4,2}$,
E. Forg\'{a}cs-Dajka$^{1}$,
J. Sztakovics$^{1,5}$and
A. B\'{o}di$^{2,3}$ 
\\\\
$^{1}$E\"{o}tv\"{o}s Lor\'{a}nd University, Department of Astronomy, H-1118 Budapest, P\'{a}zm\'{a}ny P\'{e}ter stny. 1/A, Hungary  \\
$^{2}$Konkoly Observatory, Research Centre for Astronomy and Earth Sciences, Eötvös Loránd Research Network (ELKH), H-1121 Budapest,\\ Konkoly Thege Mikl\'os \'ut 15-17, Hungary\\
$^{3}$MTA CSFK Lend\"ulet Near-Field Cosmology Research Group \\
$^{4}$Baja Astronomical Observatory of Szeged University, H-6500 Baja, Szegedi \'{u}t, Kt. 766, Hungary \\
$^{5}$Eszterh\'{a}zy K\'{a}roly University, Department of Physics, H-3300 Eger, Eszterh\'{a}zy t\'{e}r 1, Hungary 
}

\date{Accepted XXX. Received YYY; in original form ZZZ}

\pubyear{2021}

\begin{document}
\label{firstpage}
\pagerange{\pageref{firstpage}--\pageref{lastpage}}

\maketitle

\begin{abstract}

We report an eclipse timing variations (ETV) study to identify close, stellar mass companions to the eclipsing binaries monitored during the photometric survey OGLE-IV. 
We also present an alternative automatic way to determine the first and last contacts of an eclipse.
Applying the phase dispersion minimization method to identify potential triples, we find close third components with outer periods less than 1500\,days in 23 systems.  We present outer orbit solution for 21 of 23 systems. For the ten, tightest triples we find that the ETV can only be modelled with the combination of the light-travel time effect (LTTE) and third-body perturbations, while in case of another 11 systems, pure LTTE solutions are found to be satisfactory. In the remaining two systems we identify extra eclipses connected to the outer component, but for the incomplete and noisy ETV curves, we are unable to find realistic three-body solutions. Therefore, in these cases we give only the outer period.

\end{abstract}


\section{Introduction}

The analysis of hierarchical triple stellar systems plays significant role in the  understanding the formation and evolution of  short periodic binary systems \citep{Toonen2016}. 
The various formation theories of close binary systems, e.\,g. the so called Kozai Cycles with Tidal Friction (KCTF) mechanism \citep[see, e.\,g.][]{Kiselevaetal98,fabryckytremaine07,naozfabrycky14}, as well as the recently proposed different disk and core fragmentation procedures \citep{tokovinin18,moekratter18} require the presence of an additional, third  component to  explain the large number of the close, non-evolved binary stars.
 
Eclipse timing variation (ETV) analysis is a perfect method to explore hierarchical triple candidates amongst eclipsing binaries (EBs).
In this process the observed mid-times of the eclipses are compared to the pre-calculated predictions based on a constant orbital period of the system. Such a manner one get the so-called $O-C$ (Observed - Calculated) diagrams of which the shape can reveal information about the system.
When an EB has an additional companion in the system the EB's $O-C$ diagram shows a periodic variation because its distance from the observer changes as it orbits around the  center of mass of the system. This is the so called light-travel-time effect (LTTE).

In most cases, in a hierarchical triple system, the distance of the third body from the close binary's center of mass is larger than the separation of the close binary companions by orders of magnitudes, and therefore, the orbital motions of the three stars can be described with two non-perturbed Keplerian motion. However, if the system is sufficiently tight, then this approximation fails. In such systems, mutual gravitational perturbations must also be taken into account, which makes the dynamical modeling of such triples to be more complicated. On the other hand, however, their accurate modeling allows the determination of many parameters (e.g. the masses of the components) in a dynamic way \citep[see, e.g.][]{Borkovits2015}.

Thanks to the  Optical Gravitational Lensing Experiment ({\em OGLE};  \citealt{udalskietal92}), relatively long timespan\footnote{Besides the OGLE-IV data for many systems OGLE-II, III observations are also available. These three surveys together cover more than 18 years.} of light curves of around half-million eclipsing binaries (450 598 from the Galactic Bulge \citep{OGLE} and 48 605 from Large- and Small Magellanic Clouds \citep{2016ogle_eb_lmc_smc}) are available, which are suitable for discovering and studying hierarchical triple systems.

Although many similar studies have been conducted in the last decade on candidates for triple star systems associated with large surveys  (\citealt{ Borkovits2016_Kepler,Zasche_lmc,Zasche_smc,hajdu2017} and \citealt{ MOA_2018}), the influence of the dynamic effects in most of these systems, which is only characteristic of compact triple systems, is negligible. The low number of compact triple systems was also recognised by  \citet{Borkovits2016_Kepler} in the study of eclipsing binaries from the Kepler field.
According to their investigation the outer period distribution of the triple stellar systems shows a significant drop around period $P_2 < 200^d$. This phenomenon may be explained by dynamic or evolutionary processes. Another interesting thing is that systems in which the ETV is affected by significant dynamic effects are more likely to be found in this period range.

In this paper we continue our previous study \citep[][hereafter Paper I]{Hajdu2019_OGLE} to identify hierarchical triple stellar system candidates towards the Galactic Bulge with the analysis of ETVs of EBs observed during the {\em OGLE-IV} survey. Now we extend our investigations to those triple system candidates where the outer period is less than one year. 
In Paper I, instead of calculating times of minima for individual eclipses, we determined the so-called ``normal'' minima from the phase-folded average light curves of 17 consecutive binary cycles. It was done to counteract the effect of the rare sampling  of the OGLE-IV survey, which in most cases was unfavorable for the determination of individual eclipse times.
The major disadvantage of that method was, in such a manner, that the short-term ETVs were averaged out, and therefore, the tightest systems (i. e. triples with an outer to inner period ratio of, let's say $P_2/P_1 < 100$)  had likely been remained undetectable. 

In contrast to the method used in Paper I, it is necessary to determine individual minimum times in order to study really tight triple systems. This requires a method which can handle a very few (2 or 3) data points during an eclipse.
In this paper, in addition to presenting a simplified method for solving this problem, we also describe an eclipse border determination method which proved to be useful for the automatic and rapid examination of this large data set.

We formulate the basic mathematical background of the close third-body affected ETV analysis (light-travel time effect and dynamical effects) in Section \ref{effects}.
In Section \ref{basic} we outline the steps of our investigation, starting with the system selection and automatic $O-C$ curve generation, then continuing with also an automatic triple candidate selection.
The results of the ETV analysis are discussed in Section \ref{results} and here we also present some unique systems where the ETV analysis is less reliable, but extra eclipses of the third, distant component star appear several times during the observation.
Finally, a short summary is given in Section \ref{summary}.

\section{Contributions to the ETVs} \label{effects}
\subsection{Light-Travel-Time effect}
It is commonly known that most of the ETVs of hierarchical triple systems can be modeled by LTTE. The first mention of this effect can be attributed to \citet{Chandler1888} who  explained the Algol's observed ETV with this effect. The nowadays used mathematical formula of this effect was described by \citet{Irwin1952}, who also gave a graphical fitting procedure for determining the elements of the light-time orbit.

According to \citet{Irwin1952} the LTTE contribution takes the following form 
\begin{equation}
\label{LTTEfunction}
\Delta_\mathrm{LTTE}=-\frac{a_\mathrm{AB}\sin i_2}{c}\frac{\left(1-e_2^2\right)\sin\left(v_2+\omega_2\right)}{1+e_2\cos v_2},
\end{equation}
where $a_\mathrm{AB}$ denotes the semi-major axis of the EB's center of mass around the center of mass of the triple system, while $i_2$, $e_2$, $\omega_2$ stand for the inclination, eccentricity, and argument of periastron of the relative outer orbit, respectively. Furthermore, $c$ is the speed of light and $v_2$ is the true anomaly of the third component. Note the negative sign on the right hand side, which arises from the use of the {\em companion's} argument of periastron, instead of the argument of periastron of the light time orbit of the EB ($\omega_\mathrm{AB}=\omega_2+\upi$).

The amplitude of the LTTE takes the form

\begin{equation}
\label{LTTEamplitude}
\mathcal{A}_\mathrm{LTTE}=\frac{a_\mathrm{AB}\sin i_2}{c}\sqrt{1-e_2^2\cos^2\omega_2},
\end{equation}
while the mass function $f(m_\mathrm{C})$, analogous to its spectroscopic counterpart for single-line spectroscopic binaries, is usually defined as
\begin{equation}
\label{Mass_function_eq}
f(m_\mathrm{C})=\frac{m_C^3\sin^3i_2}{m^2_\mathrm{ABC}}=\frac{4\pi^2a^3_\mathrm{AB}\sin^3i_2}{GP_2^2},
\end{equation}
where $ m_C $ is the mass of the third body, $ m_ {ABC}$ is the mass of the system and G is the gravitational constant.   It can be calculated from the parameters of the LTTE solution.  
Note that if the mass of the eclipsing binary is known, the minimum mass ($i_2=90^\circ$) of the third component can be determined.

\subsection{Dynamical effects}

In tight hierarchical triple stellar systems the perturbations of the third body may also modify the ETV significantly within a short time scale. The analytic formula that describes this dynamical ETV was described in detail in a series of papers by \citet{Borkovits2003,Borkovits2011,Borkovits2015}. Therefore, here we only present the main relations between the orbital parameters and the $O-C$ diagrams.

The dynamical ETV component ($\Delta_\mathrm{dyn}$) not only depends on the orbital elements of the inner and outer orbit but also on their relative configurations. Furthermore the relative orientation to the observer is also an additional important factor if the inner orbit is eccentric. 
A comprehensive description of these effects can be found in \cite{Borkovits2015}.
In our sample, however, we calculate dynamical effects only for EBs with small eccentricity. Therefore, for our purposes it is satisfactory to use the substantially simpler formula of \citet{Borkovits2003}, which assumes a  circular inner orbit, as follows:

\begin{eqnarray}
\Delta_\mathrm{dyn}&=&\frac{3}{4\pi}\frac{m_C}{m_{ABC}}\frac{P_1^2}{P_2}(1-e_2^2)^{-3/2} \nonumber\\
& &\times\left[\left(\frac{2}{3}-\sin^2i_\mathrm{m}\right)\mathcal{M}+\frac{1}{2}\sin^2i_\mathrm{m}\mathcal{S}\right],
\label{Eq:Deltadynlead}
\end{eqnarray}
where

\begin{eqnarray}
\mathcal{M}&=&v_2-l_2+e_2\sin v_2, \label{eq_M}\\
\mathcal{S}&=&\sin(2v_2+2g_2)+e_2 \nonumber \\ 
&&\times\left[\sin(v_2+2g_2)+\frac{1}{3}\sin(3v_2+2g_2)\right] \label{eq_S}.
\end{eqnarray}
The first part of Eq. \ref{Eq:Deltadynlead} is an amplitude-like quantity that depends on the mass ($m_C/m_{ABC}$) and period ($P_1^2/P_2$) ratios and the eccentricity of the outer orbit ($e_2$). In  the second part of the equation $\mathcal{M}$ and $\mathcal{S}$ stand for the time-dependent functions of the true anomaly ($v_2$) as well as the mean anomaly ($l_2$) of the outer body on its relative orbit around the EB's  center of mass, while $g_2$ is the tertiary's {\em dynamical} argument of periastron measured from the intersection of the inner and outer orbital planes.

\section{Basic steps of the analysis} \label{basic}
\subsection{System preselection} 
In order to identify tight triples with short outer periods, in contrast to the method used in Paper I,  we attempted to determine accurate mid-minima times of individual eclipses. Naturally, it needs better sampled light curves.
Therefore, we selected those light curves which contain more than 4000 data points in I band.
Applying  this criteria we reduced our sample from the original $\sim 450\,000$  EBs to $\sim80\,000$ systems.

\subsection{Determination of times of minima}
\label{determ_time_of_minima}

To determine the times of minima of the pre-selected sample, the first step is to create an eclipse model curve for each star. To do so, we used phase-folded and binned light curves (FBLC).
For the folding, the 
periods were taken from the OGLE catalog\footnote{\url{http://www.astrouw.edu.pl/ogle/ogle4/OCVS/blg/ecl/ecl.dat}}. 
The folded light curves were binned into $N$ equally spaced phase-cells, according to the orbital phases of each measured points where the value of $N$ depends on the orbital period of the binary system in the following way:
\begin{itemize} 
\item $N = 1000$ if the orbital period is shorter than 5 days, 
\item $N = 2000$ if the orbital period is between 5 and 10 days,
\item $N = 5000$ if the orbital period is longer than 10 days. 
\end{itemize}
Then the median flux were calculated cell by cell. Note that if the phase curve is not fully covered than we used a simple linear function to fill these phase bins. 

In this paper we used polynomial template functions derived from the FBLC for the primary and secondary eclipses to determine the mid-times of the individual eclipse events.
The generation of eclipse template functions also requires the phase boundaries of eclipses, which need to be selected quickly and automatically when examining a large sample of EBs similar to OGLE. 
In case of detached and semi-detached systems the determination of the eclipse borders is quite challenging due the various shapes of the FBLCs. For this purpose we have developed a simple but effective method as follows  (demonstrated in Fig 1.).

\begin{figure*}
    \includegraphics[width=2.\columnwidth]{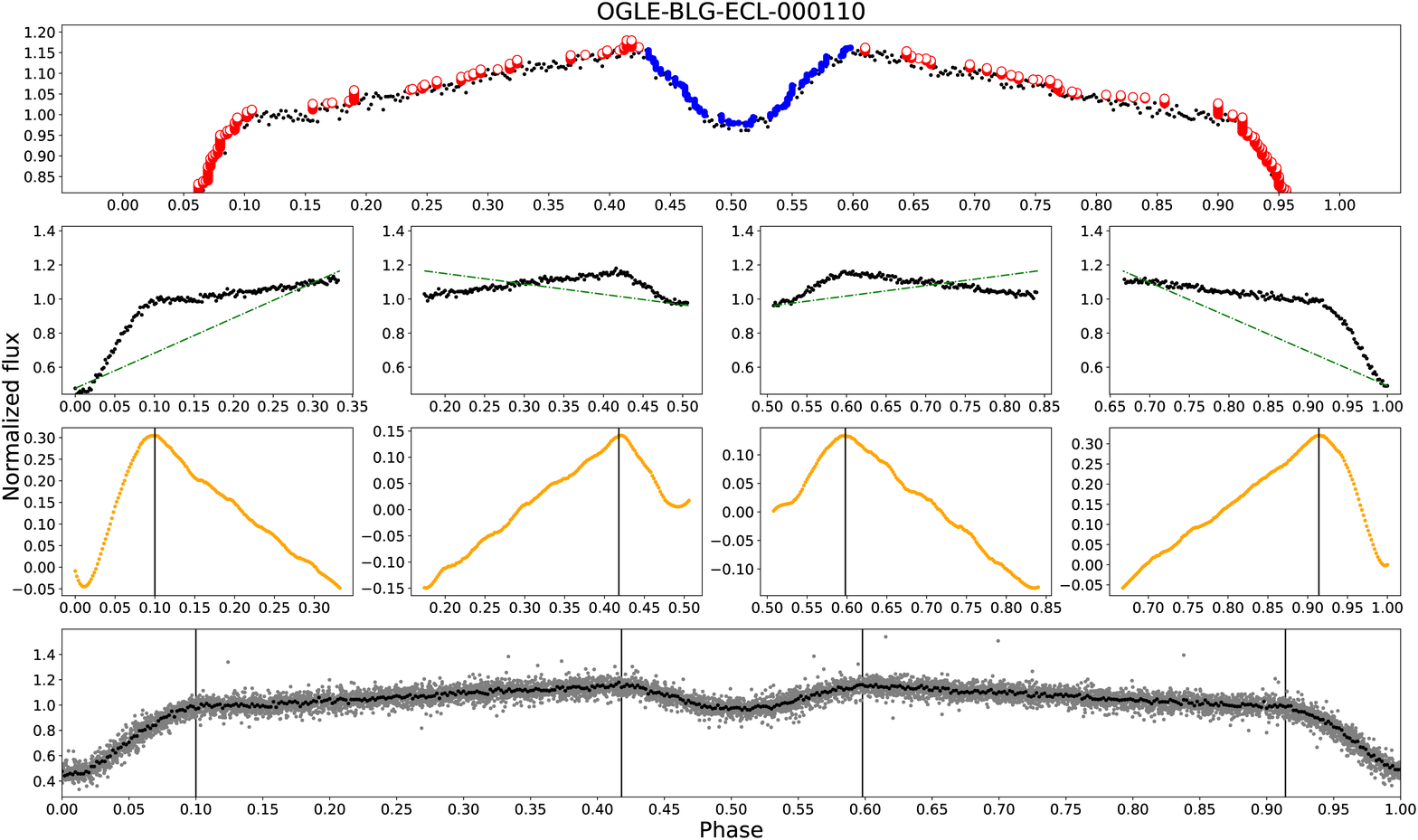}
    \caption{
    The workflow of the minimum determining algorithm. The first panel represents the possible eclipse borders for all different flux levels on the FBLC (first step). Note these points (red and blue) are just the possible eclipse borders if we use flux levels for eclipse border estimation. The red points belong to the primary, the blue ones belong to the secondary eclipse and the black points represents the FBLC. 
    The panels in the second line show the FBLC (black) and the lines which were used for subtraction and which were described in  Section \ref{determ_time_of_minima} (dashed green lines). The panels in the third line show the subtracted and smoothed FBLCs (orange) which maxima are good estimations for eclipse borders. The phase values of the resulting eclipse borders are indicated by the vertical solid black lines.  
    Finally, the bottom panel shows the primary and the secondary eclipse borders, along with the original phase curve (gray points) and the FBLC.}
    \label{fig:borders}
\end{figure*}

Assuming that the primary eclipses are located around photometric phase $\phi=0$, our first task is to find the locations of the secondary eclipses. For the search of the secondary eclipse the code departs from the flux level corresponds to the average of the minimum and maximum fluxes. The program determines the borders of those phase intervals where most of the points have remained under this level. Then this flux level is increased with small steps. Thousand steps are needed to reach the maximum flux value, however, the code stops if the secondary minimum point determined in 50 consecutive steps, have remained unchanged. Then this phase value is considered to be the (approximate) minimum phase of the secondary eclipse.
Note that this process was used only to determine the approximate phase of the secondary eclipse.
This method was found to be effective even if due to the strong reflection effects the secondary eclipse is higher than the average flux of the system (see the uppermost panel of Fig.~\ref{fig:borders}). Furthermore, it was working also well in case of noisy FBLCs with shallow secondary eclipses.

The next task is to find borders of the eclipses, practically the locations of the first and last contacts which are used for the determination of the individual minima times (see below). Now we illustrate the process looking for the right border (last contact) of primary eclipse. In order to find this point the code takes the straight line that connects the points ($\phi_\mathrm{primin}$;$\ell_\mathrm{primin}$) and ($\phi=\phi_\mathrm{primin}+0.5$;$\ell_\mathrm{max}$), where $\ell_\mathrm{primin}$ and $\ell_\mathrm{max}$ stand for flux at the mid-point of the primary eclipse (i.e. at phase $\phi_\mathrm{primin}$), and the maximum flux value of the whole FBLC, respectively. Then, this straight line is subtracted from the FBLC. Finally, the phase of the maximum point of the obtained residual curve is considered as the location of the last contact point, as it is shown in the middle left panels of Fig.~\ref{fig:borders}. The other three contact points are identified in similar manner (see the other panels of Fig.~\ref{fig:borders}).

In the next step, the program generates 8-th order polynomial template functions for both eclipses (primary and secondary). The minimum point of the template functions for the primary and the secondary eclipses were calculated by Newton's method using the points with the lowest flux value from the appropriate part of the  FBLC as an initial guess for the iterations.

To determine the mid-times of the individual eclipses, the program used Levenberg-Marquardt algorithm to fit the eclipses using the template functions. 
Because of the low sampling ratio and the short outer period we fixed all the parameters in the function we fit but the \textit{shifting} parameter ($\phi'$):

\begin{equation}\label{elc_eq}
    g(\phi) = a_0 + a_1 \cdot f(\phi+\phi'),
\end{equation}
where $a_0$ and $a_1$ are the additive and multiplicative parameters respectively, $f$ is the template function mentioned above and $\phi'$ stands for the the phase shift of the eclipse in question. This latter parameter stands for the  observed delay of the mid-time of the eclipse in phase.
In our case, we have to suppose that the brightness of the components and the eclipse depth do not change considerably by time. For this reason, we checked each hierarchical triple system candidate LC.

\subsection{Search for triple candidates}
For reliable and automatic search for short periodic variations of ETVs, we applied phase dispersion minimization (PDM; \citealt{Stellingwerf1978_PDM}), which is a well-known technique to determine the period of variable stars. 
With this method, we searched for the main periods of $O-C$s between 50 and 1000 days.
Because searching for the period is more appropriate in the frequency space we used the frequency mode of the \textit{pyPDM} class of the \textit{PyAstronomy}\footnote{\url{https://github.com/sczesla/PyAstronomy}} package  \citep{pya} with 0.0001 frequency steps.

After applying this method to the primary and the secondary $O-C$ diagrams separately, we selected those systems where the two periods found by the PDM were close to each other.
We considered two periods to be close if their difference was less than 5 days or less than the 5\% of the period of the primary. 

However, in addition to LTTE and dynamical effects, several other effects can cause periodic signals in the O-C curves, making it difficult to select triple stellar candidates. 
Good examples are the stellar spots which can significantly affect the ETV curves \citep{Balaji_Kepler_spots}, therefore the program automatically found these, too. 
However, in case of spotted stars the secondary eclipse varies in opposite phase, therefore, they can be easily filtered out by fitting sinusoidal curves to both (primary and secondary) folded $O-C$s and comparing their phases. If they are different, the most likely scenario is spots cause the periodic change.

\begin{figure}
    \centering
    \includegraphics[width=\columnwidth]{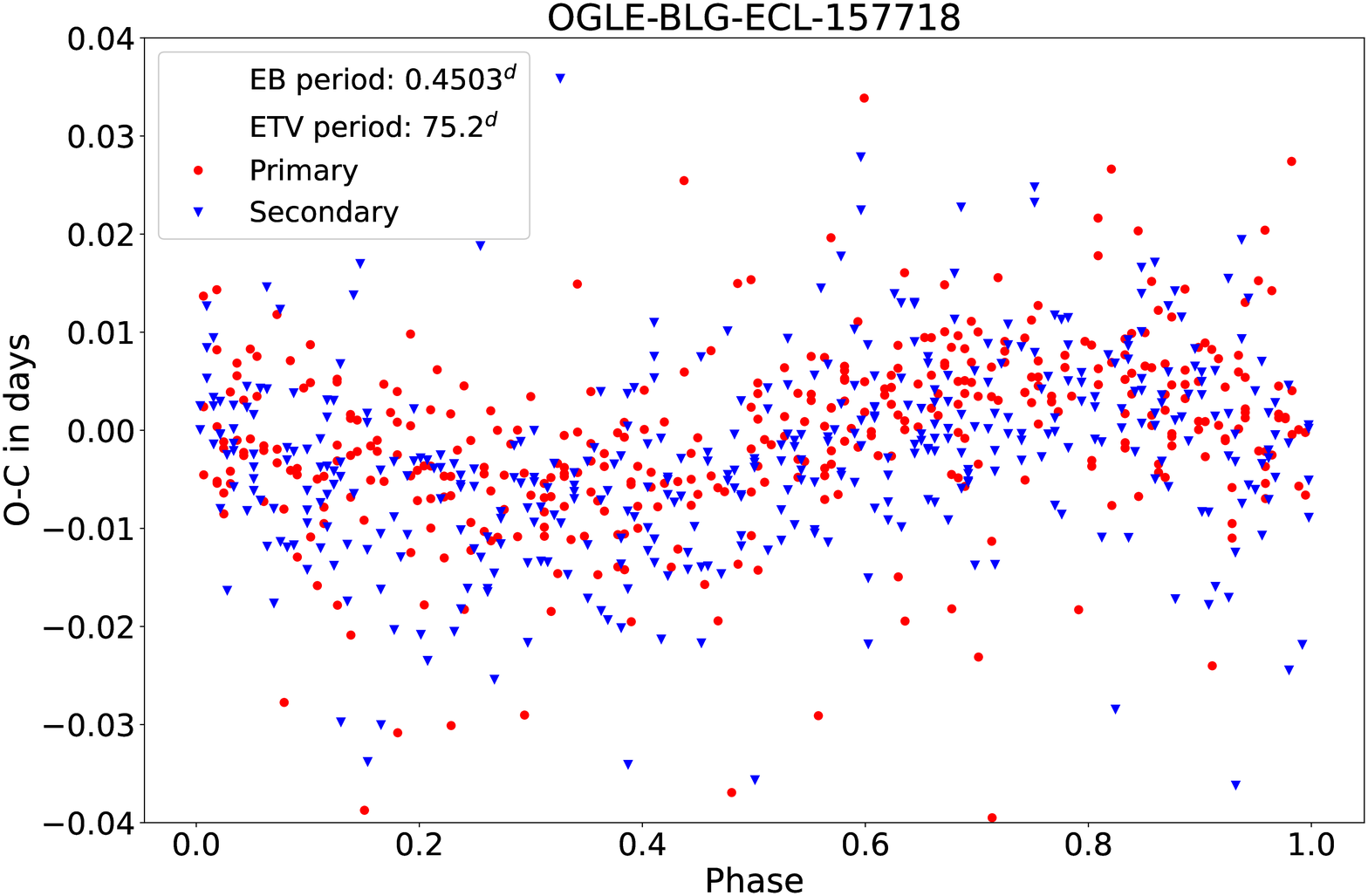}
    
    \includegraphics[width=\columnwidth]{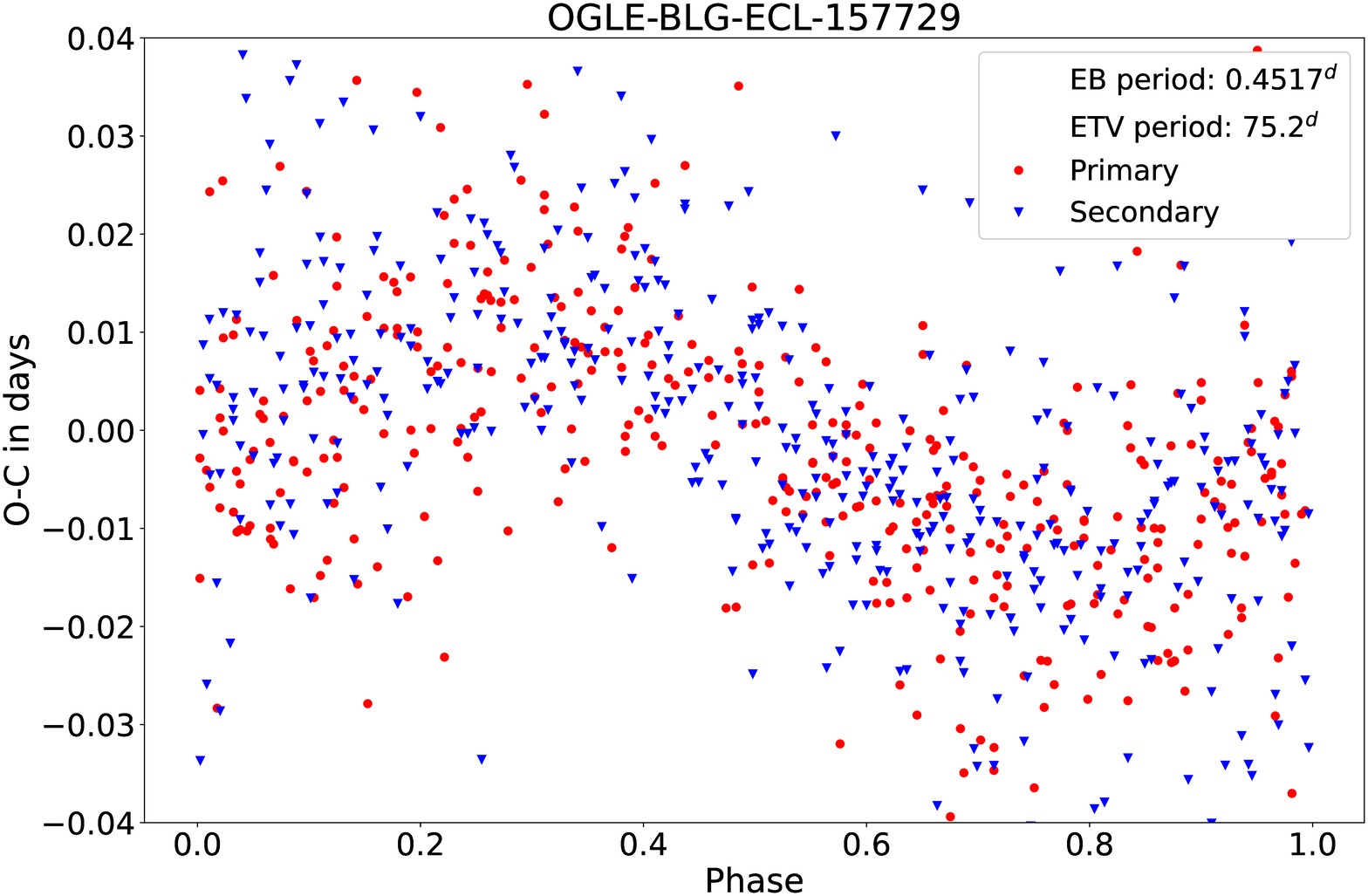}
    \caption{Folded $O-C$s of OGLE-BLG-ECL-157718 (top) and OGLE-BLG-ECL-157729 (bottom). These systems  are blended binaries and this fact caused that their ETVs show a false periodic trend with the same period.}
    \label{fig:blended}
\end{figure}

Another problem is the blended variables when two close periods cause a beat. Note there is no physical connection between the two eclipsing binary. These stars can have significant influence on the binary's LC thus getting periodic $O-C$s. A good example for this effect can be seen in Fig. \ref{fig:blended} in the case of OGLE-BLG-ECL-157718 and OGLE-BLG-ECL-157729, where both show ETVs with periods of around 75 days. However, they are blending each other which cause misleading sinusoidal features in their ETVs. Actually the half of the difference between the orbital frequencies appears on the $O-C$ curves of the systems.

The final list of the short periodic triple candidates were selected manually from the candidates provided by the automatic program to avoid false-positive candidates, where the seasonal visibility period dominates the PDM spectrum. In some cases the program found potential new hierarchical systems with true outer period longer than $1000^d$.
The 21 new hierarchical triple candidates 
are listed in Table \ref{tab_LTTE_dyn} and \ref{tab_only_LTTE}.

\subsection{Fits and error estimation of minimum times}
 We found that the parameters in Eq.~\ref{Mass_function_eq} can easily converge to unphysical values during the fit. Because of this, instead of the $a\cdot\sin(i_2)$ and $m_c/m_{abc}$, we used the parameters $m_{ab}\cdot\sin^3(i_2)$  and $m_c\cdot\sin^3(i_2)$ in the formulas, with bounds of $0-2M_\odot$ and $0-5M_\odot$, respectively. After the fit, we transformed these formulas and their parameters back to their original form. 

For the error estimations of the minimum times of the candidates we used Markov chain Monte Carlo (MCMC)  method (as implemented in the \textit{emcee}\footnote{\url{https://emcee.readthedocs.io/en/stable/}} package; \citealt{emcee}) with the usual 3 parameter fit ($a_0$, $a_1$ and $\phi'$ parameters in Eq. \ref{elc_eq}) where it was possible. 
In cases of the systems where the lack of the points did not allow this method (less than 3 points/eclipse), we fitted only the \textit{shifting} parameter ($\phi'$). For each eclipse, the following settings were used :
\begin{itemize}
    \item number of walkers ($nwalkers$) $= 10$,
    \item number of iteration ($niter$) $=5000$ and
    \item the first half of the samples was discarded and the second half was used for the error estimation.
\end{itemize}

The parameter uncertainties are the 16-th and 84-th percentiles of the posterior distributions.

\section{Results} 	\label{results}
As a result we found 21 new short periodic hierarchical triple stellar systems and two triply eclipsing candidates.
The orbital parameters and their fitted values are presented in Table \ref{tab_LTTE_dyn} for systems with both LTTE and dynamical effects, and in Table \ref{tab_only_LTTE} for systems with pure LTTE solutions. The ETVs of the candidates with the best model fits and the folded $O-C$s are shown in Appendix \ref{Appendix_A} and \ref{Appendix_B}.

\subsection{Hierarchical triples with significant dynamical effects}
\label{res_dyn_candidates}
Based on the periods ($P_1$ and $P_2$) we can deduce whether the dynamic term should also be taken into account in addition to the LTTE. 
Basically, if the period of the outer orbit ($P_2$) is less than the outer period of a hypothetical system with dynamical amplitude $\mathcal{A}_{dyn} = 50^s$ and for which the orbital parameters are as follows:
\begin{itemize}
    \item $P_1$ = period of the EB,
    \item $e_2 = 0.35$,
    \item $i_2 = 60^\circ$,
    \item $\omega_2 = +/- 90^\circ$
\end{itemize}
then the parameters of the dynamical effects were also fitted. Note that this is the same boundary used by \citet{Borkovits2016_Kepler}.

In this way we found 10 candidates with measurable dynamical effects.
The values of the fitted parameters ($P_2$, $a_{AB}\sin(i_2)$, $e_2$, $\omega_2$, $\tau_2$, $i_m$, $g_2$, $m_C/m_{ABC}$) are listed in Table \ref{tab_LTTE_dyn}. In this table we also present the ratio of the amplitudes of the dynamical and LTTE effects. 
In addition, where a phase shift between the $O-C$s of the primary and the secondary minima was well visible we fit this difference as a function of eccentricity ($e_1$) and the argument of the periastron ($\omega_1$) of the inner orbit. This shift took the following form:
\begin{equation}
\Delta_{sec-pri}=\frac{ 2}{\pi}\cdot e_1\cdot\cos(\omega_1)\cdot P_1.
\end{equation}

The same MCMC algorithm was used to fit the parameters of Eq. \eqref{LTTEfunction}. and \eqref{Eq:Deltadynlead} with $nwalks=500$,  $niter=30000$ and the first half of the steps were considered to be part of the burn-in phase. We used the the parameters of the model with the smallest $\chi^2$
and their most reliable errors were calculated based on the 16-th and the 84-th percentiles. 
In this section we only present just a few of the systems in more detail.

\begin{figure}
    \centering
    \includegraphics[width=\columnwidth]{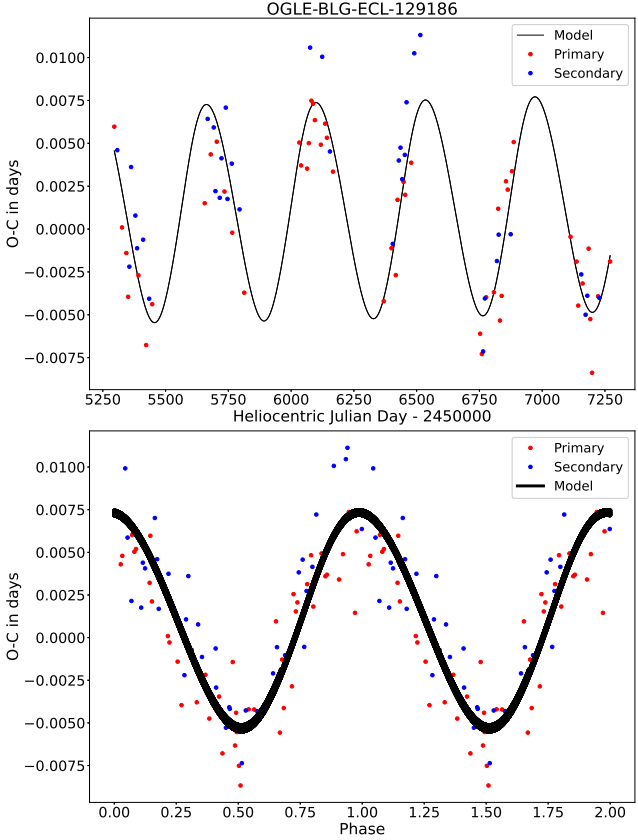}
    \caption{The ETV of OGLE-BLG-ECL-129186 triple system candidate ($P_1=3.4284091^d$ and $P_2\approx432.6^d$) with our model (top panel), and the folded $O-C$ of the system after we removed the linear trend (bottom panel).  
 } 
    \label{fig:129186_oc}
\end{figure}

\textit{OGLE-BLG-ECL-129186} is the perfect example to demonstrate the opportunities in the PDM analysis of ETV because the well separated parts can easily mislead the naked eye which thus only appear to be random noise (see Fig. \ref{fig:129186_oc}). Note that this system has longer outer period ($P_2\approx 430^d$), but because of the dynamical effects we listed it here.    

\begin{figure}
    \centering
    \includegraphics[width=\columnwidth, trim={0 25cm 0 0},clip]{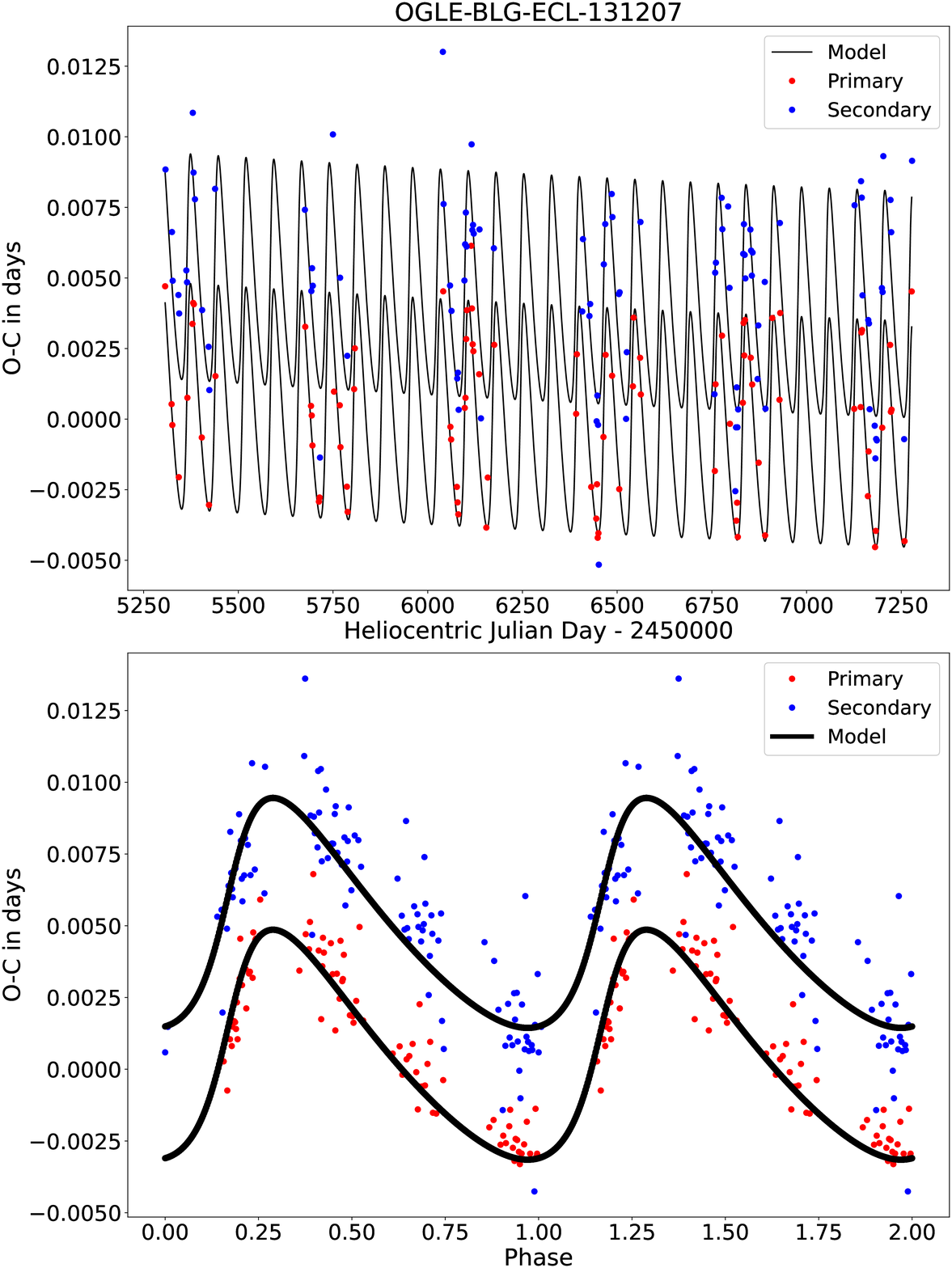}
    \caption{The ETV of OGLE-BLG-ECL-131207 triple system candidate ($P_{1} = 1.8967999^d$ and $P_2\approx 73.3^d$) with our model. Almost two outer cycles are visible within each observational season. The small vertical difference between the primary and the secondary minima is due to the small inner eccentricity. 
 } 
    \label{fig:131207_oc}
\end{figure}

\textit{OGLE-BLG-ECL-131207} is a slightly eccentric Algol-type eclipsing binary, that has the shortest external period ($P_2\approx 73.3^d$) found so far among OGLE-IV triple systems. 
As a result, almost two full outer periods are clearly visible within each observational season in its ETV as it is shown in Fig. \ref{fig:131207_oc}.

\begin{figure}
    \centering
    \includegraphics[width=\columnwidth,height=0.63\columnwidth]{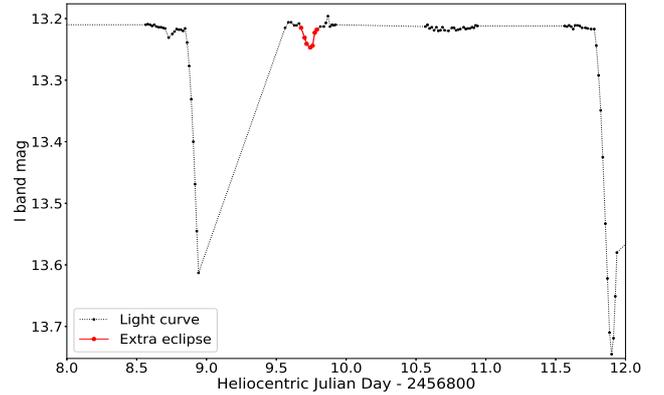}
    \caption{Light curve of \textit{OGLE-BLG-ECL-161428} ($P_{1} = 2.9271820^d$ and $P_2\approx154.6^d$) shows a possible third eclipse around HJD 2456809.7. Its appearance time agrees with the predicted time for this third eclipse based on our solution. Its depth is almost 0.04 magnitude, which is 
    10 times larger than the magnitude error. }
    \label{fig:extra_event}
\end{figure}

\textit{OGLE-BLG-ECL-161428} is the only system where we were able to identify a third eclipse between HJD 2456809.67 and 2456809.80 (Fig. \ref{fig:extra_event}) which coincides with its expected time. Thus, we can say that this extra eclipse in the LC is probably caused by the third companion. However, based on the LC, this is the only extra eclipse, caused by the third body, which was observed. We only accepted those parts of the LC where more than 2 data points were lower than 3 standard deviations from the average light curve - here the eclipse depth ($\approx0.04$ mag) is almost 10 times larger than the magnitude error.

\begin{figure}
    \centering
    \includegraphics[width=\columnwidth, trim={0 25cm 0 0},clip]{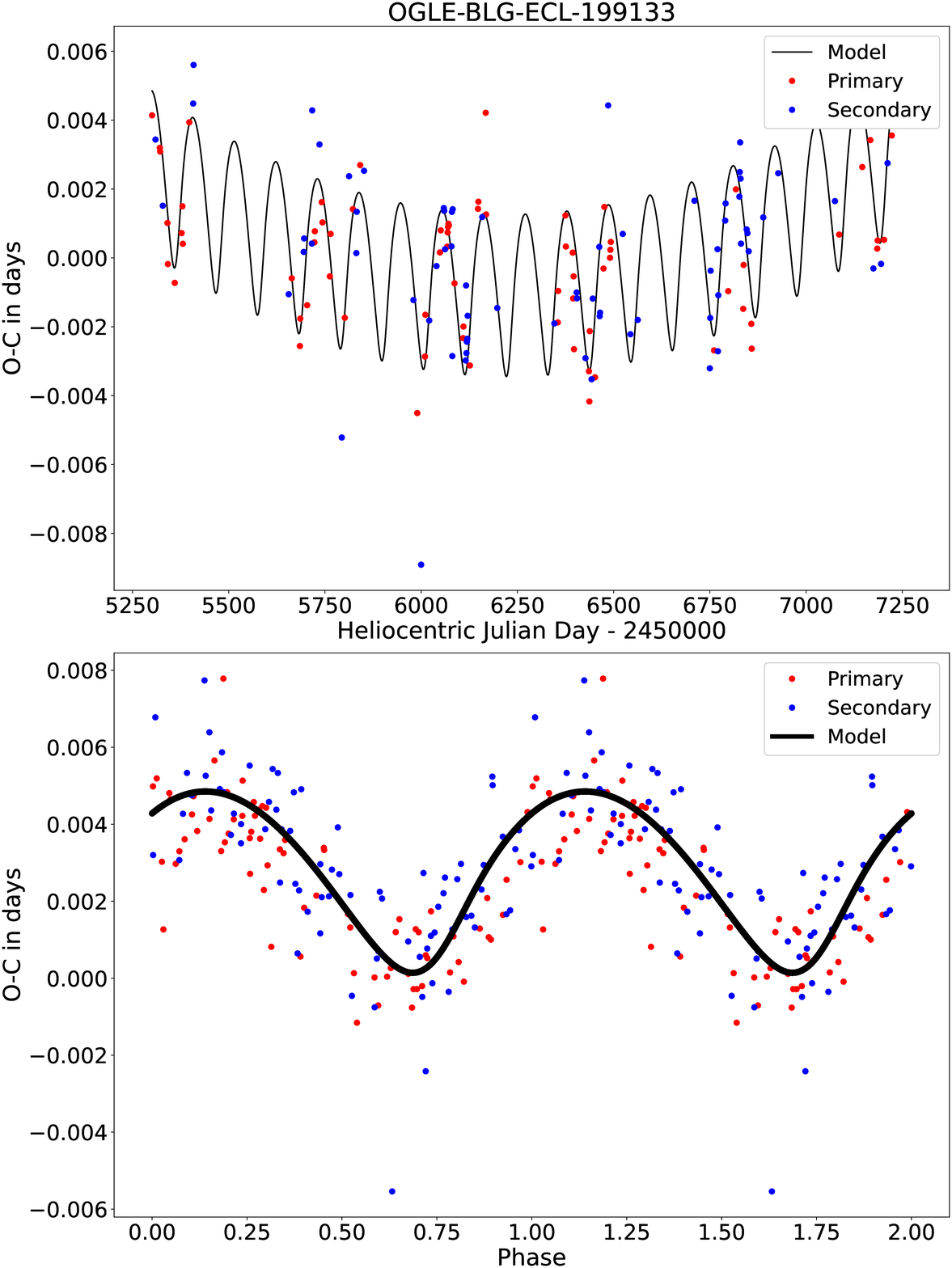}
    \caption{ETV of OGLE-BLG-ECL-199133 ($P_{1} = 1.0518475^d$ and $P_2\approx107.8^d$) which beside the sinusoidal variation shows a significant parabolic trend.
    }
    \label{fig:199133_oc}
\end{figure}

\begin{figure}
    \centering
    \includegraphics[width=\columnwidth]{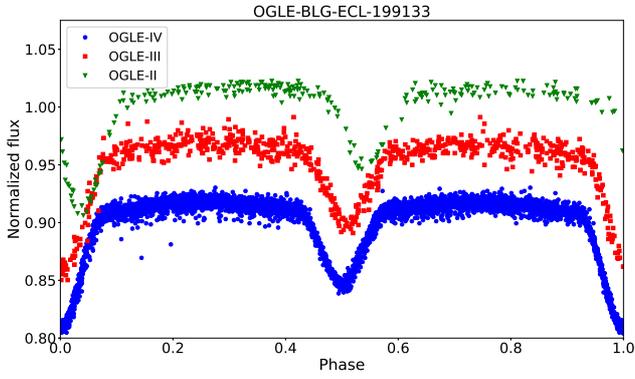}
    \caption{Folded light curves of OGLE-BLG-ECL-199133 from OGLE-II,-III and -IV which show a slight horizontal shift because of the increasing period value. Note, for these folded light curves we used the period, corrected based on its OGLE-IV ETV (Fig \ref{fig:199133_oc}). For the better visibility the different OGLE series were shifted vertically. }
    \label{fig:199133_folded}
\end{figure}

\textit{OGLE-BLG-ECL-199133} is an Algol-type binary with period around 1 day and its eccentricity is really close to zero. It is a bright system compared to most of the candidate systems since its average brightness is $\sim13.25^m$ in I band. Its ETV (Fig. \ref{fig:199133_oc}) shows a significant parabolic trend which suggests mass transfer between the components, similar to VW~Cep and DX~Cyg, which were analysed by \citet{Mitnyan2018} and recently comprehensively studied by \citet{DX_Cygni}, respectively. This statement is also supported by the fact that the folded light curves of the system show a well visible shift plotted in Fig. \ref{fig:199133_folded}. For these light curves we used the corrected period of the system based on its ETV.
Based on our analyses the parabolic trend seen in the $O-C$ of \textit{OGLE-BLG-ECL-199133} can be explained by an increasing period value with a period change rate of $\dot{P}\sim0.26\pm0.02$ seconds per year.

\begin{figure}
    \centering
    \includegraphics[width=\columnwidth]{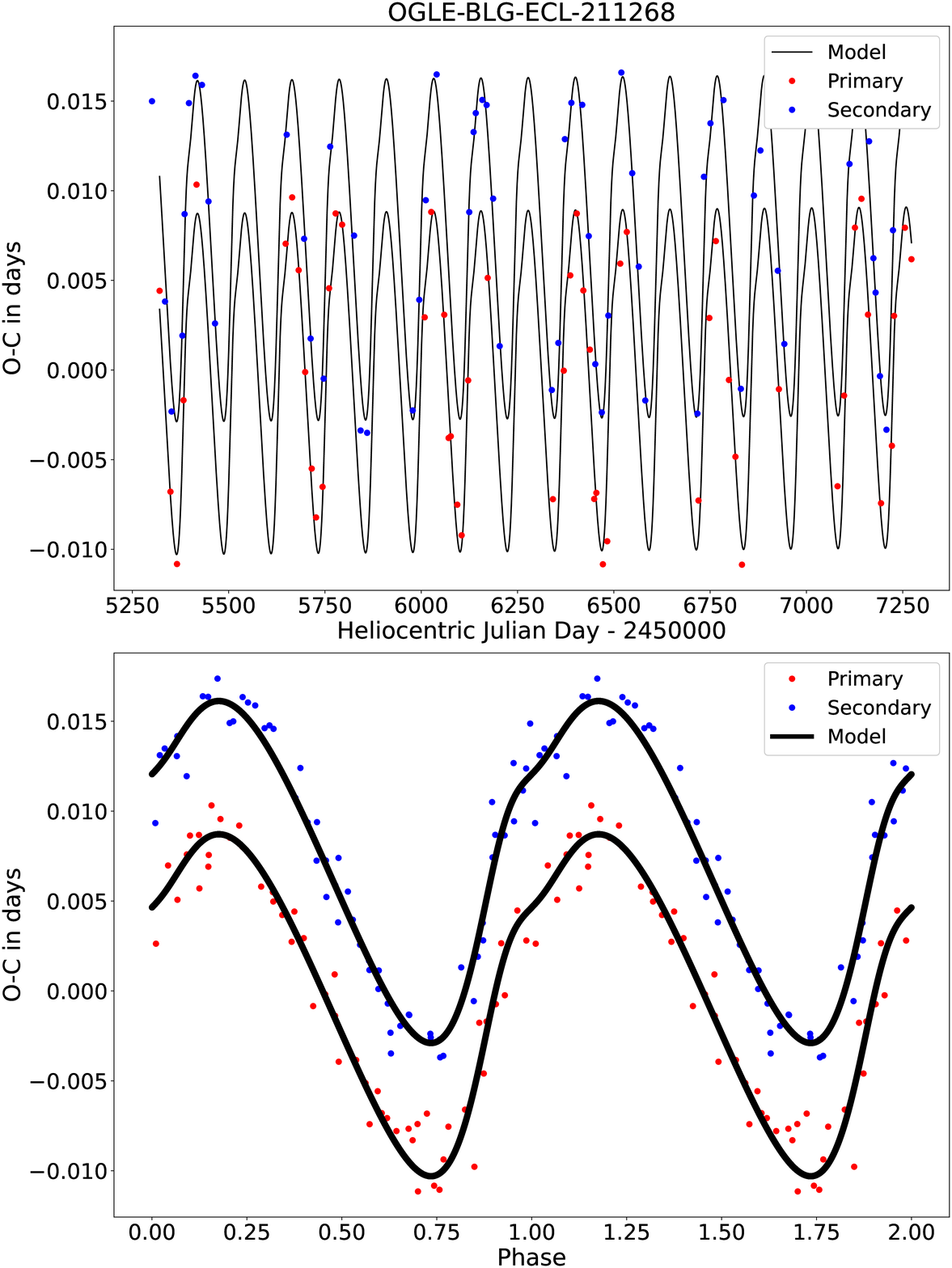}
    \caption{ETV of OGLE-BLG-ECL-211268 (top panel) and the folded $O-C$ of the system (bottom panel). The $O-C$ curves of the system carry the signs of the dynamical effects which can be easily recognised around phase value $\approx 1$, which is also due to the small $P_2/P_2$ ratio ($P_1 =5.6414329^d$ and $P_2\approx122.7^d$).   }
    \label{fig:211268o-c}
\end{figure}

\textit{OGLE-BLG-ECL-211268}'s light curve has the highest signal to noise ratio and therefore its O-C (Fig. \ref{fig:211268o-c}) and the fitted orbital parameters have the lowest relative uncertainties. Its uniqueness is also manifested in the fact that the influence of the dynamic effect can be clearly seen in the folded $O-C$ curve as well (bottom panel of Fig. \ref{fig:211268o-c}).
Furthermore, it has the longest EB period of the triple system candidates, even having the second longest period if we take into account the candidates from Paper I.

However, \textit{OGLE-BLG-ECL-253744} is the third brightest star from our candidates the eclipse depth of the primary minimum is only 0.05 magnitude. 
Its outer-to-inner period ratio ($P_2/P_1$) is the lowest from our list, therefore, the dynamical effects have to be the dominant. Note that only simplified mathematical formulae were  used during our investigations, therefore the parameters we received are not relevant except the period of the outer orbit.

\begin{landscape}
\begin{table}
    \centering
    \caption{Orbital elements from combined dynamical and LTTE solutions. Note that the last candidate is separated because of the most likely false values of the parameters.}
    \label{tab_LTTE_dyn}
    \begin{tabular}{|ccccccccccccccc|}
\hline
ID & $P_1$ & $P_2$ &  $a_\mathrm{AB}\sin(i)$ & $e_2$ & $\omega_2$ & $\tau_2$ & $i_m$ & $g_2$ & $m_C/m_{ABC}$  & $e_1\cos(\omega_1)$ & $\frac{\mathcal{A}_\mathrm{dyn}}{\mathcal{A}_\mathrm{LTTE}}$\\

 & [days] & [days]  & [$R_{\odot}$] &  & [rad] & [days] & [rad] & [rad]  &  &  & \\ \hline
 

    129186  &  3.4284091  &  436.03 $^{ +7.33 }_{ -7.33 }$ &  201.79 $^{ +33.37 }_{ -36.39 }$ &  0.16 $^{ +0.14 }_{ -0.11 }$ &  2.84 $^{ +1.42 }_{ -1.37 }$ &  5542.59 $^{ +82.36 }_{ -97.14 }$ &  0.41 $^{ +0.28 }_{ -0.28 }$ &  1.53 $^{ +0.96 }_{ -0.90 }$ &  0.59 $^{ +0.12 }_{ -0.09 }$ & - & 0.21 \\ [0.2cm] 


    131207  &  1.8967999  &  73.32 $^{ +0.04 }_{ -0.05 }$ &  43.83 $^{ +27.87 }_{ -16.80 }$ &  0.35 $^{ +0.09 }_{ -0.09 }$ &  -1.17 $^{ +1.00 }_{ -1.66 }$ &  5365.12 $^{ +1.53 }_{ -1.94 }$ &  0.24 $^{ +0.19 }_{ -0.16 }$ &  3.11 $^{ +2.16 }_{ -2.09 }$ &  0.46 $^{ +0.21 }_{ -0.15 }$ & 0.0038 $^{ +0.0002 }_{ -0.0002 }$ &  3.61\\[0.2cm]


    144579  &  2.1691786  &  77.45 $^{ +0.11 }_{ -0.13 }$ &  41.05 $^{ +45.17 }_{ -22.09 }$ &  0.31 $^{ +0.14 }_{ -0.11 }$ &  2.35 $^{ +2.94 }_{ -1.73 }$ &  5356.01 $^{ +3.88 }_{ -3.31 }$ &  0.31 $^{ +0.16 }_{ -0.20 }$ &  3.08 $^{ +2.23 }_{ -2.27 }$ &  0.42 $^{ +0.29 }_{ -0.21 }$  & - & 3.54\\[0.2cm]


    151751  &  1.1756323  &  99.99 $^{ +0.13 }_{ -0.14 }$ &  49.89 $^{ +15.85 }_{ -12.59 }$ &  0.37 $^{ +0.09 }_{ -0.13 }$ &  2.29 $^{ +0.60 }_{ -0.52 }$ &  5393.56 $^{ +4.49 }_{ -7.70 }$ &  0.50 $^{ +0.22 }_{ -0.32 }$ &  3.15 $^{ +2.16 }_{ -1.99 }$ &  0.44 $^{ +0.10 }_{ -0.10 }$ & - & 0.74\\[0.2cm]
        

    161428  &  2.9271820  &  154.59 $^{ +0.32 }_{ -0.32 }$ &  57.72 $^{ +50.28 }_{ -23.93 }$ &  0.21 $^{ +0.06 }_{ -0.10 }$ &  0.77 $^{ +0.98 }_{ -0.49 }$ &  5366.19 $^{ +8.36 }_{ -10.37 }$ &  0.22 $^{ +0.14 }_{ -0.15 }$ &  3.17 $^{ +2.02 }_{ -2.21 }$ &  0.38 $^{ +0.26 }_{ -0.15 }$ & 0.0002 $^{ +0.0001 }_{ -0.0001 }$ & 1.37\\[0.2cm]


    199133  &  1.0518475  &  107.85 $^{ +0.49 }_{ -0.48 }$ &  77.25 $^{ +19.39 }_{ -14.62 }$ &  0.20 $^{ +0.18 }_{ -0.14 }$ &  2.19 $^{ +1.92 }_{ -1.63 }$ &  5368.44 $^{ +29.61 }_{ -27.76 }$ &  0.48 $^{ +0.43 }_{ -0.33 }$ &  3.08 $^{ +2.03 }_{ -2.17 }$ &  0.58 $^{ +0.15 }_{ -0.10 }$ & - & 0.25 \\[0.2cm]


    211268  &  5.6414329  &  122.67 $^{ +0.03 }_{ -0.03 }$ &  44.22 $^{ +3.97 }_{ -4.31 }$ &  0.28 $^{ +0.01 }_{ -0.02 }$ &  4.77 $^{ +0.36 }_{ -0.38 }$ &  5391.36 $^{ +0.37 }_{ -0.42 }$ &  0.45 $^{ +0.02 }_{ -0.02 }$ &  4.38 $^{ +0.03 }_{ -0.03 }$ &  0.32 $^{ +0.02 }_{ -0.02 }$ & 0.0020 $^{ +0.0001 }_{ -0.0001 }$ & 8.10\\[0.2cm]


    217867  &  5.4566043  &  168.89 $^{ +0.57 }_{ -0.53 }$ &  104.78 $^{ +62.31 }_{ -51.03 }$ &  0.26 $^{ +0.11 }_{ -0.07 }$ &  0.62 $^{ +1.49 }_{ -2.34 }$ &  5385.84 $^{ +5.83 }_{ -5.31 }$ &  0.27 $^{ +0.14 }_{ -0.18 }$ &  3.60 $^{ +0.60 }_{ -0.87 }$ &  0.58 $^{ +0.21 }_{ -0.24 }$ & - & 4.59\\[0.2cm]


    249742  &  5.3456048  &  204.86 $^{ +0.96 }_{ -1.97 }$ &  57.77 $^{ +72.83 }_{ -35.14 }$ &  0.03 $^{ +0.11 }_{ -0.00 }$ &  0.93 $^{ +0.80 }_{ -1.82 }$ &  5375.12 $^{ +92.29 }_{ -3.51 }$ &  0.21 $^{ +0.06 }_{ -0.17 }$ &  2.23 $^{ +2.81 }_{ -1.28 }$ &  0.32 $^{ +0.31 }_{ -0.19 }$ & - & 0.38\\[0.2cm]

\hline 
    
    253744  &  5.5073876  &  82.72 $^{ +0.08 }_{ -0.12 }$ &  40.63 $^{ +51.34 }_{ -7.91 }$ &  0.44 $^{ +0.03 }_{ -0.15 }$ &  1.31 $^{ +4.87 }_{ -0.22 }$ &  5385.20 $^{ +2.18 }_{ -0.73 }$ &  0.23 $^{ +0.14 }_{ -0.15 }$ &  2.69 $^{ +2.87 }_{ -1.89 }$ &  0.38 $^{ +0.34 }_{ -0.04 }$ & 0.0145 $^{ +0.0008 }_{ -0.0008 }$ & 34.5\\[0.2cm]

    \end{tabular}
\end{table}

\begin{table}
    \centering
    \caption{Orbital elements from LTTE solutions.
    }
    \label{tab_only_LTTE}
    \begin{tabular}{c  c c c c c c c }
\hline
ID & $P_1$ & $P_2$ & $a_\mathrm{AB}\cdot\sin(i_2)$ & $e_2$ & $\tau_2$ & $\omega_2$ & $f(m_\mathrm{C})$ \\
 & [days] & [days]  & [$R_{\odot}$] &  & [days] & [rad] & \\ \hline

    135829  &  1.9585061 &  542.90 $^{+ 2.01 }_{- 2.24 }$ &  119.50 $^{+ 3.42 }_{- 3.08 }$ &  0.18 $^{+ 0.05 }_{- 0.06 }$ &  5788.04 $^{+ 34.29 }_{- 20.22 }$ &  4.52 $^{+ 0.36 }_{- 0.30 }$ & 0.08 $^{+ 0.01 }_{- 0.01 }$ \\[0.2cm]
    
    158649  & 1.5779421  &  322.49 $^{+ 3.42 }_{- 4.02 }$ &  71.08 $^{+ 14.02 }_{- 6.93 }$ &  0.50 $^{+ 0.27 }_{- 0.17 }$ &  5543.75 $^{+ 26.82 }_{- 19.64 }$ &  5.34 $^{+ 0.34 }_{- 0.39 }$ & 0.05 $^{+ 0.02 }_{- 0.03 }$ \\[0.2cm]
    
    169013  & 0.4531956  &  485.45 $^{+ 5.58 }_{- 6.67 }$ &  161.88 $^{+ 18.81 }_{- 12.97 }$ &  0.25 $^{+ 0.23 }_{- 0.14 }$ &  5797.49 $^{+ 69.29 }_{- 31.89 }$ &  6.86 $^{+ 0.97 }_{- 0.36 }$ & 0.25 $^{+ 0.07 }_{- 0.09 }$ \\[0.2cm]
    
    172213  & 0.4673387  &  1064.07 $^{+ 4.88 }_{- 4.55 }$ &  165.28 $^{+ 70.58 }_{- 13.79 }$ &  0.93 $^{+ 0.02 }_{- 0.05 }$ &  5796.70 $^{+ 5.09 }_{- 3.29 }$ &  3.05 $^{+ 0.03 }_{- 0.04 }$ & 0.07 $^{+ 0.03 }_{- 0.07 }$ \\[0.2cm]
    
    180010  & 1.5445121  &  1195.96 $^{+ 15.00 }_{- 9.92 }$ &  243.36 $^{+ 15.70 }_{- 9.46 }$ &  0.58 $^{+ 0.06 }_{- 0.06 }$ &  5321.52 $^{+ 16.77 }_{- 15.71 }$ &  2.56 $^{+ 0.13 }_{- 0.09 }$ & 0.14 $^{+ 0.02 }_{- 0.02 }$ \\[0.2cm]
    
    181440  & 0.9522017  &  320.23 $^{+ 2.33 }_{- 1.78 }$ &  98.87 $^{+ 16.41 }_{- 9.94 }$ &  0.45 $^{+ 0.18 }_{- 0.11 }$ &  5255.98 $^{+ 19.52 }_{- 20.56 }$ &  2.17 $^{+ 0.24 }_{- 0.41 }$ & 0.14 $^{+ 0.04 }_{- 0.08 }$ \\[0.2cm]

    185642  & 0.4274139  &  1045.62 $^{+ 16.75 }_{- 18.31 }$ &  173.71 $^{+ 20.11 }_{- 11.96 }$ &  0.29 $^{+ 0.13 }_{- 0.13 }$ &  6128.13 $^{+ 90.48 }_{- 76.31 }$ &  6.40 $^{+ 0.57 }_{- 0.48 }$ & 0.07 $^{+ 0.01 }_{- 0.02 }$ \\[0.2cm]

    189164  &  1.3383325 &  751.60 $^{+ 28.09 }_{- 23.55 }$ &  159.92 $^{+ 36.99 }_{- 10.06 }$ &  0.41 $^{+ 0.23 }_{- 0.21 }$ &  5959.00 $^{+ 56.89 }_{- 110.80 }$ &  0.55 $^{+ 0.89 }_{- 0.45 }$ & 0.11 $^{+ 0.03 }_{- 0.06 }$ \\[0.2cm]

    221167  & 0.3180689  &  876.01 $^{+ 14.86 }_{- 19.68 }$ &  217.07 $^{+ 28.42 }_{- 18.10 }$ &  0.44 $^{+ 0.17 }_{- 0.17 }$ &  5540.12 $^{+ 70.57 }_{- 58.51 }$ &  3.95 $^{+ 0.48 }_{- 0.45 }$ & 0.18 $^{+ 0.05 }_{- 0.07 }$ \\[0.2cm]
    
    280301  &  0.3980531 &  1246.67 $^{+ 29.49 }_{- 20.41 }$ &  265.03 $^{+ 36.25 }_{- 18.57 }$ &  0.38 $^{+ 0.17 }_{- 0.15 }$ &  6222.04 $^{+ 44.81 }_{- 81.21 }$ &  1.67 $^{+ 0.32 }_{- 0.33 }$ & 0.17 $^{+ 0.04 }_{- 0.07 }$ \\[0.2cm]
    
    294058  & 0.3076762  &  1025.47 $^{+ 21.90 }_{- 34.70 }$ &  131.20 $^{+ 10.32 }_{- 8.06 }$ &  0.42 $^{+ 0.10 }_{- 0.25 }$ &  6116.53 $^{+ 93.55 }_{- 51.07 }$ &  3.93 $^{+ 0.77 }_{- 0.20 }$ & 0.03 $^{+ 0.01 }_{- 0.01 }$ \\

\hline
    \end{tabular}
\end{table}

\end{landscape}

\subsection{Triples with only LTTE solution}

We also found 11 new triple system candidates which were undetected by our previous method described in Paper I and which outer periods are shorter than 1500 days. The orbital parameters of these systems were also calculated via MCMC method but since their ETV can be well described by the LTTE, which is much simpler to fit, therefore the number of iterations was reduced to $niter=1000$. The parameters of these systems are presented in Table \ref{tab_only_LTTE}. Here we only used LTTE solution since $\mathcal{A}_\mathrm{dyn} < 50^s $ (see Fig. \ref{fig:per-per}). 

However, there are two systems, where $\mathcal{A}_\mathrm{dyn} \approx 50^s$, therefore we estimated the amplitude of the dynamic term based on the following equation:
\begin{equation}
\mathcal{A}_\mathrm{dyn}=\frac{1}{2\upi}\frac{m_\mathrm{C}}{m_\mathrm{ABC}}\frac{P_1^2}{P_2}\left(1-e_2^2\right)^{-3/2},
\label{Eq:dyn_amp}
\end{equation}
where the mass of the components were taken as 1 solar mass and the period ($P_2$) and eccentricity ($e_2$) were based on the LTTE solution. We found that the dynamical term in both systems is negligible.

\subsection{Triple candidates exhibits outer eclipses}

Triply eclipsing systems are quite rarely observed among hierarchical triples not only because it needs a special configuration, but because it also needs a long term and continuous observation. Despite these facts the number of known triply eclipsing hierarchical systems is slowly rising \citep{Lee2013,hajdu2017,Borkovits2019,Mitnyan2020}.

Thanks to the manual inspection we also found some systems where automatic searching algorithm failed. This may have been due to the sampling frequency (missing observations of primary or secondary minima), or the low-quality $O-C$ diagram of one of the components, where no periodic variability was identified.

\begin{figure}
    \centering
\includegraphics[width=\columnwidth]{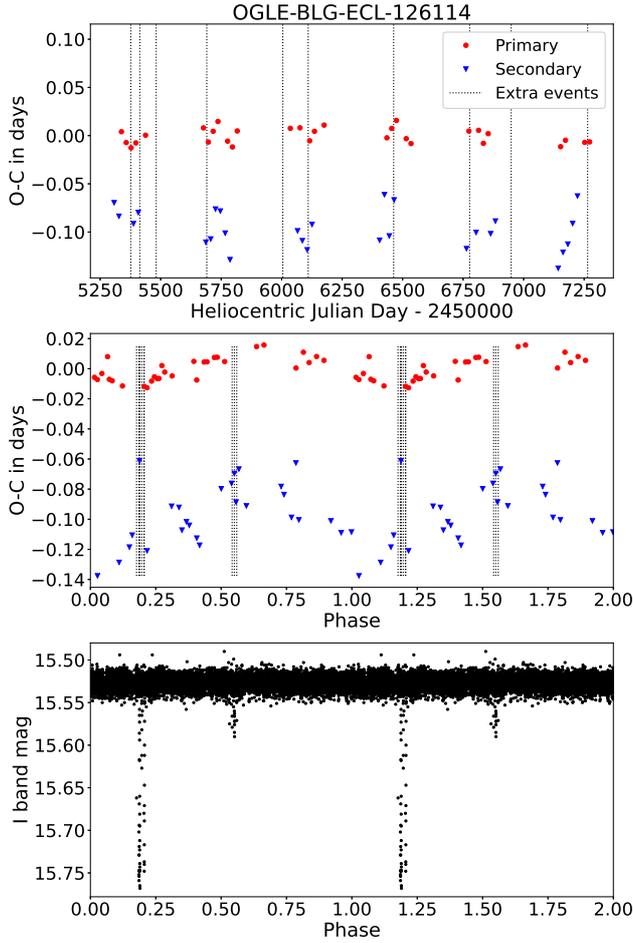}
    \caption{The top panel shows the ETV of OGLE-BLG-ECL-126114 ($P_1=10.1261144^d$. The black vertical lines indicate times where further eclipses have appeared. The phase folded ETVs and the phase values of the extra eclipses are shown in the middle panel. Lower panel represents the folded LC of the system ($P_2 = 105^d$) without the eclipses of the close binary.} 
    \label{fig:126114_oc}
\end{figure}

\textit{OGLE-BLG-ECL-126114} is an Algol type binary with a relatively shallow secondary minimum. The average eclipse depth of the secondary minima is $\approx 0.015$ magnitude, therefore the $O-C$s of the secondary minima is of low quality. Because of this, the period analysis of the $O-C$ curves of the primary and the secondary data resulted in different $P_2$ values. However, the ETV of the primary minima shows clear periodicity with $P_2 =105^d$. The folded LC of the system also shows extra eclipse-like features. After the primary and secondary eclipses were cut out from the original LC, the remaining data showed a periodicity with a period equals to the $P_2$ that was derived from the primary's ETV (see in Fig. \ref{fig:126114_oc}). Based on these we conclude that the extra eclipses belong to this hierarchical system, which outer period is $P_2\approx 105^d$, instead of belonging to a blended EB, as it was noted by the OGLE team\footnote{\url{ftp://ftp.astrouw.edu.pl/ogle/ogle4/OCVS/blg/ecl/remarks.txt}}. However, further observations are needed to determine the outer orbital parameters of this system.

\begin{figure}
    \centering
    \includegraphics[width=\columnwidth]{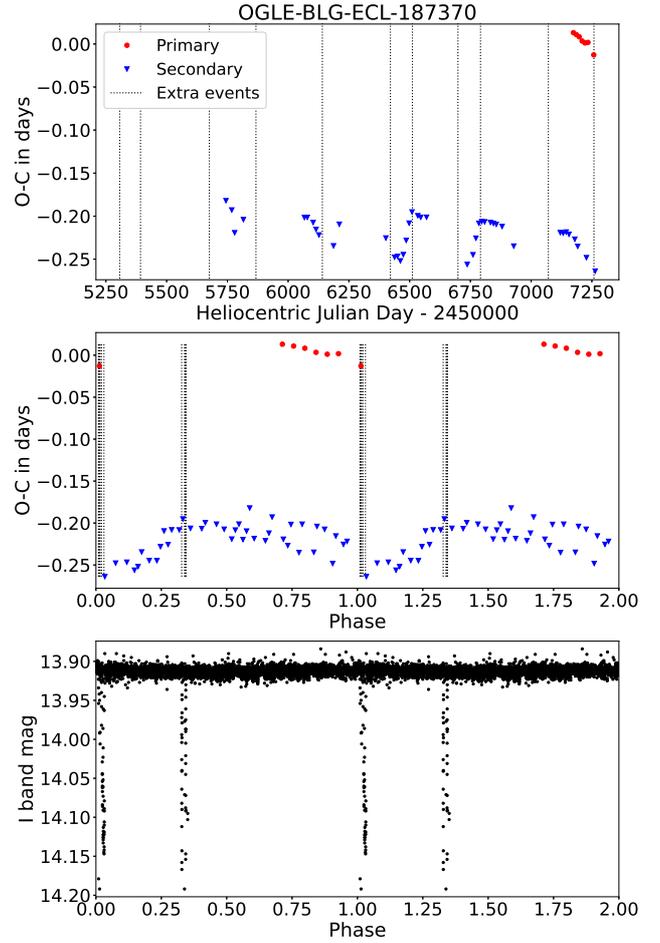}
    \caption{The top panel shows the ETV of OGLE-BLG-ECL-187370 ($P_1=11.9634963^d$. The black vertical lines indicate times where further eclipses appeared. It is well visible that the secondary minima have a periodic variation and the extra eclipses appears around the same phases. The phase folded ETVs and the phases value of the extra eclipses are shown in the middle panel. Lower panel represents the folded LC of the system ($P_2 = 279^d$) without the eclipses of the close binary. } 
    \label{fig:187370_oc}
\end{figure}

\textit{OGLE-BLG-ECL-187370} is another system which was found manually by checking the automatically generated $O-C$ diagrams. The ETV of this system is plotted in Fig. \ref{fig:187370_oc}.
Ground-based observations are not conducive to observe systems with orbital period close to an integer (measured in days) because the folded light curves are thus incomplete. In our case, due to the almost 12-day period ($P_2=11.9635$), only a small part of the primary minimum was observed.
The most interesting features of this system are the additional clear and deep eclipses which were considered to be eclipses of a second, blended binary with period of $\approx81$ days \citep{OGLE}. The probable quadruple nature of this target was investigated by \citet{Zasche_doble_ecl}, but they found no evidence for that. 
In fact, the ``blended'' system's orbital period is $279^d$ which is almost equal to the orbital period we got from the PDM analysis of the system's ETV in question. This fact may point to an assumption that this is another triply eclipsing system.

\subsection{Period-period distribution}
The period-period ($P_1-P_2$) distribution of the hierarchical triple candidates is plotted in Fig. \ref{fig:per-per}. Here, following Figure 8 from \citet{Borkovits2016_Kepler}, we present the $50^s$ amplitude borders for LTTE and dynamical effects (blue lines). These limits were calculated for a hypothetical triple system of three, equally solar mass stars, with a typical outer eccentricity of $e_2=0.35$, and quite arbitrarily, $i_2=60\degr$ and $\omega_2=\pm90\degr$.
We also show the position of the line of the dynamical stability (oblique red line) based on the  hierarchical triples in \cite{2001MNRAS.321..398M}. We also plotted the approximate period limit of ordinary contact binaries ($P_1=0.2^d$; vertical red line). For comparison we plotted the known \textit{OGLE-IV} triple system candidates (adopted from \citealt{Hajdu2019_OGLE}) toward the galactic bulge (gray points).

From the distribution of our candidates it can be seen that the automatic method was able to find tight systems with period ratios ($P_2/P_1$) lower than 100. However, for systems with inner periods longer than 6 days, we were unable to detect any signs of a third body with this method. This could be explained by the low number of eclipses in case of EBs with long orbital periods. For most binaries with orbital period of $\sim5^d$ less than 70 primary minima are covered at least by 2 observations, which is the minimum required to fit individual eclipses.
Moreover, based on the recently published data of \cite{Bodi2021}, the period distribution of most likely Algol-type OGLE bulge binaries (that have morphology parameter $c \le 0.4$), which dominate the $P_2>5^d$ region, have a peak around $P_1\sim 6^d$, which is followed by a significant, monotonic decrease towards longer periods. This is the second reason for not being able to find more hierarchical triple systems.

\begin{figure}
    \centering
    \includegraphics[width=\columnwidth]{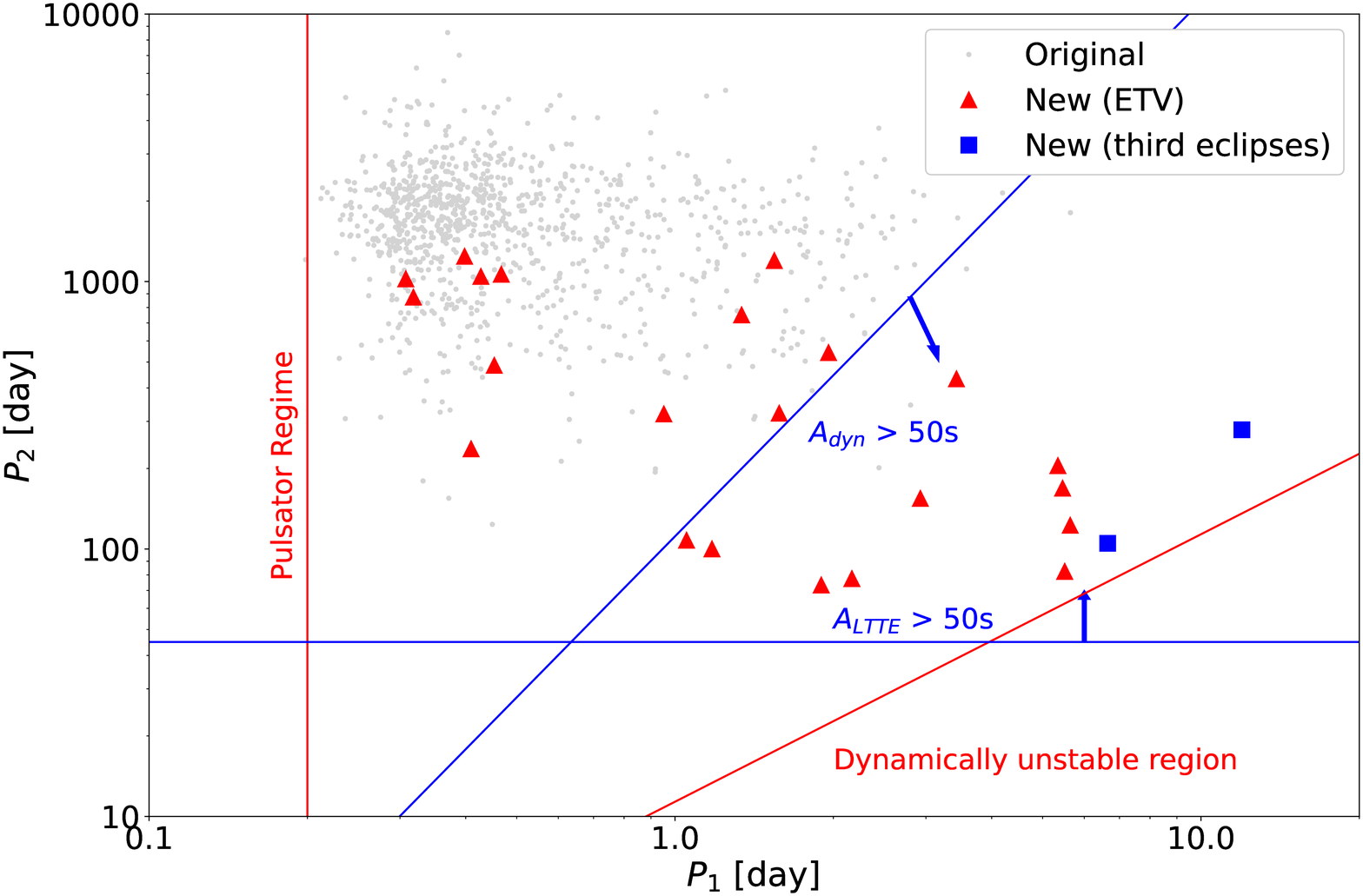}
    \caption{The location of the 21 new triple star candidates (red triangles) in the $P_1$ vs $P_2$ plane. We also added the triple eclipsing systems to the figure (blue squares). For comparison we plotted the known \textit{OGLE-IV} triple system candidates (adopted from \citealt{Hajdu2019_OGLE}) toward the galactic bulge. These systems are marked with gray points.
    The horizontal and the oblique blue lines show the borders of the domains where the amplitudes of the LTTE and dynamical terms may exceed $\sim50\,$sec, respectively, which can be regarded as a limit for an unambiguous detection.
    The region under the rising red line is the dynamically unstable region, in the sense of the stability criteria of \citet{mardlingaarseth01}.}
    \label{fig:per-per}
\end{figure}

\section{Summary and conclusion} \label{summary}

In this paper we present an alternative method to automatically produce and analyse the ETV of EBs in the interest of searching for short periodic hierarchical triples or even spotted stars in EB systems.

As a result we found 21 new hierarchical triple stellar candidates in the Galactic bulge, of which 10 shows significant dynamical effects; to fit the ETVs of the other 11 systems LTTE solution were satisfactory. Besides, we also present two triple eclipsing hierarchical triple stellar systems where the primary and the secondary eclipse of the outer orbit is clearly visible, but the ETV analysis alone does not provide sufficient information about the orbital parameters because of the LC's low quality or the unfortunate EB orbital period, which is if close to the an integer (measured in days) makes the observation of eclipses not suitable for ground-based studies.

Our main results are summarized in Fig. \ref{fig:per-per}, where we show the period-period distribution of the candidates, including the two systems, which were classified as triple candidates based only on the third eclipses, and the candidates from Paper I. Out of the 21 new systems, 10 show not negligible dynamical effects, however, the parameter values, especially from the dynamical term, are not reliable. Photodynamical analysis is needed for better results which would be published in a further paper.

After our investigation there is no doubt that OGLE is still a ''\textit{treasure chest}'' of multiple stellar systems. This is also supported by the fact that the more recently published paper on double eclipsing binaries by \cite{Zasche_doble_ecl} also presents several new candidates.

\section*{Acknowledgements}
This project has been supported by the Hungarian Research, Development and Innovation Office, NKFIH-OTKA grants
KH-130372 and the KKP-137523 `SeismoLab` \'Elvonal grant, the Lend\"ulet Program, project No. LP2018-7/2020, the MW-Gaia COST Action (CA18104). Also the authors are grateful to Anik\'{o} Farkas-Tak\'{a}cs, R\'obert Szab\'o and Roz\'alia \'Ad\'am for their comments and helps.

\section*{Data Availability}

The photometry data for OGLE-IV Eclipsing Binaries toward the Galactic Bulge is publicly available and can be downloaded from \url{http://www.astrouw.edu.pl/ogle/ogle4/OCVS/blg/ecl/}. The derived data generated in this research will be shared on reasonable request to the corresponding author.



\bibliographystyle{mnras}
\bibliography{ogle2} 




\appendix

\section{ETV of hierarchical triple stellar systems with dynamical effects}
\label{Appendix_A}

\begin{figure*}
    \centering
    \includegraphics[width=\columnwidth]{Figs_new/129186_O-C_all.eps}
    \includegraphics[width=\columnwidth]{Figs_new/131207_O-C_all.eps}
    
    \includegraphics[width=\columnwidth]{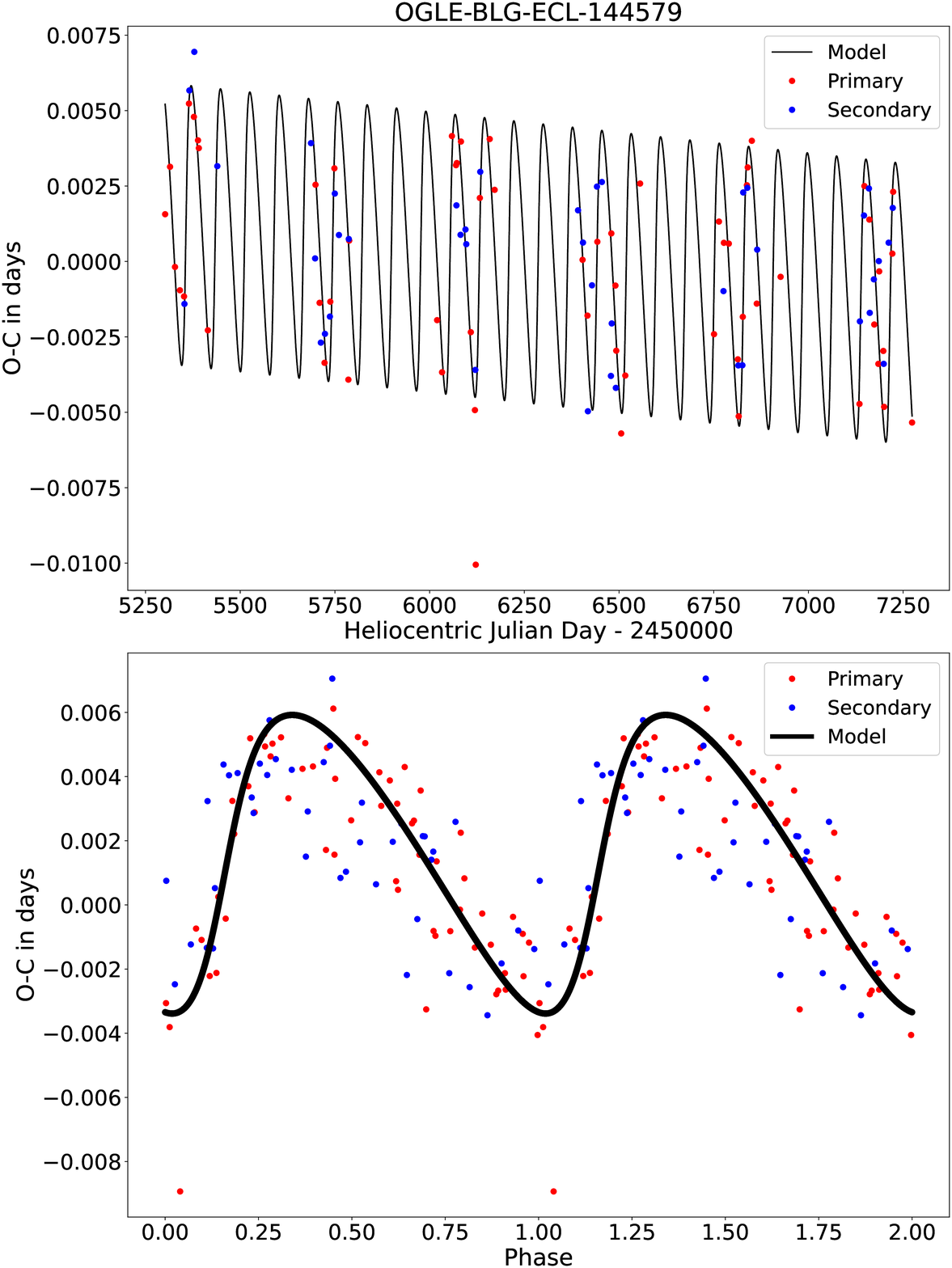}
    \includegraphics[width=\columnwidth]{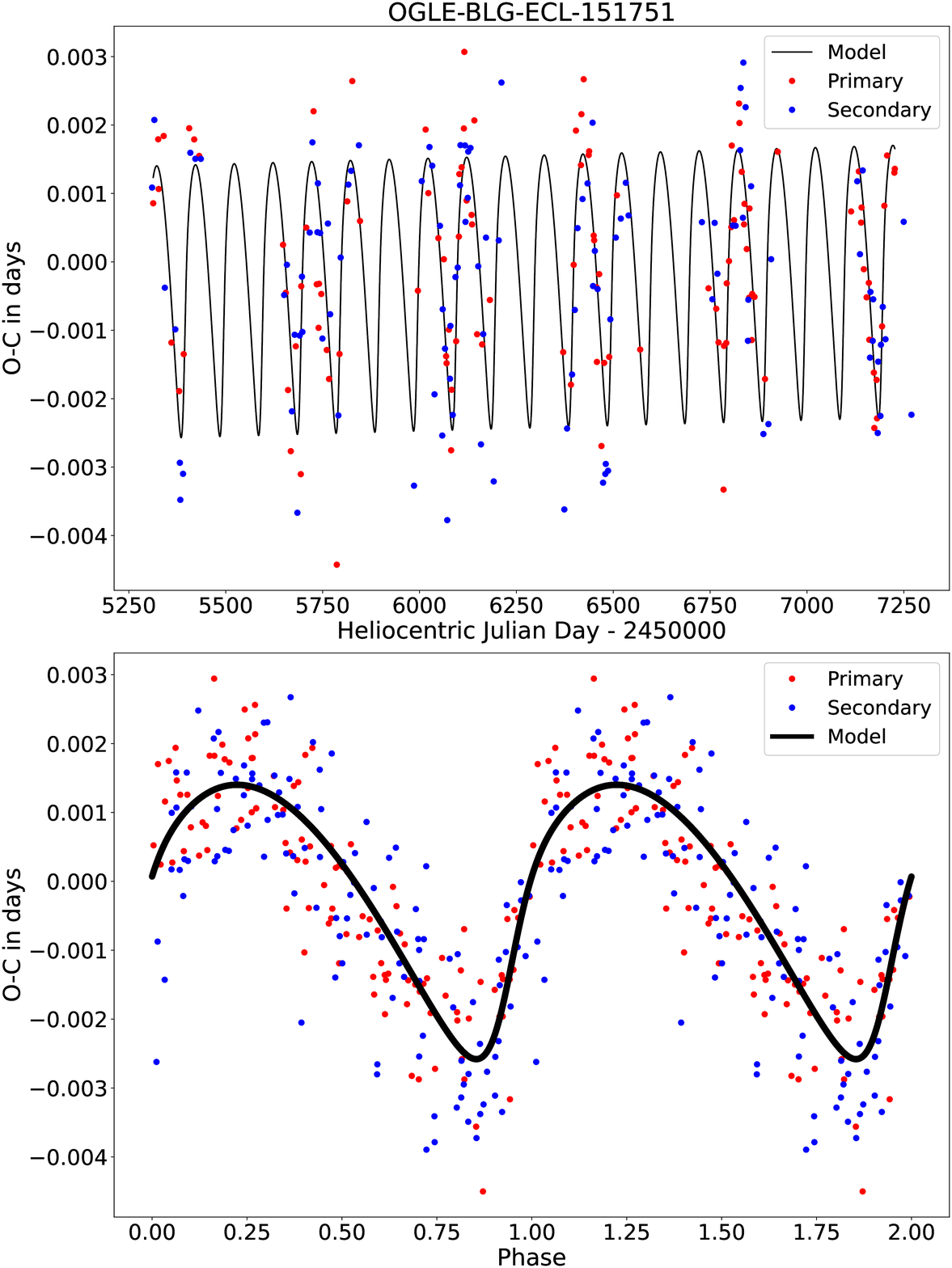}
    \clearpage
\end{figure*}

\begin{figure*}
    \centering
    \includegraphics[width=\columnwidth]{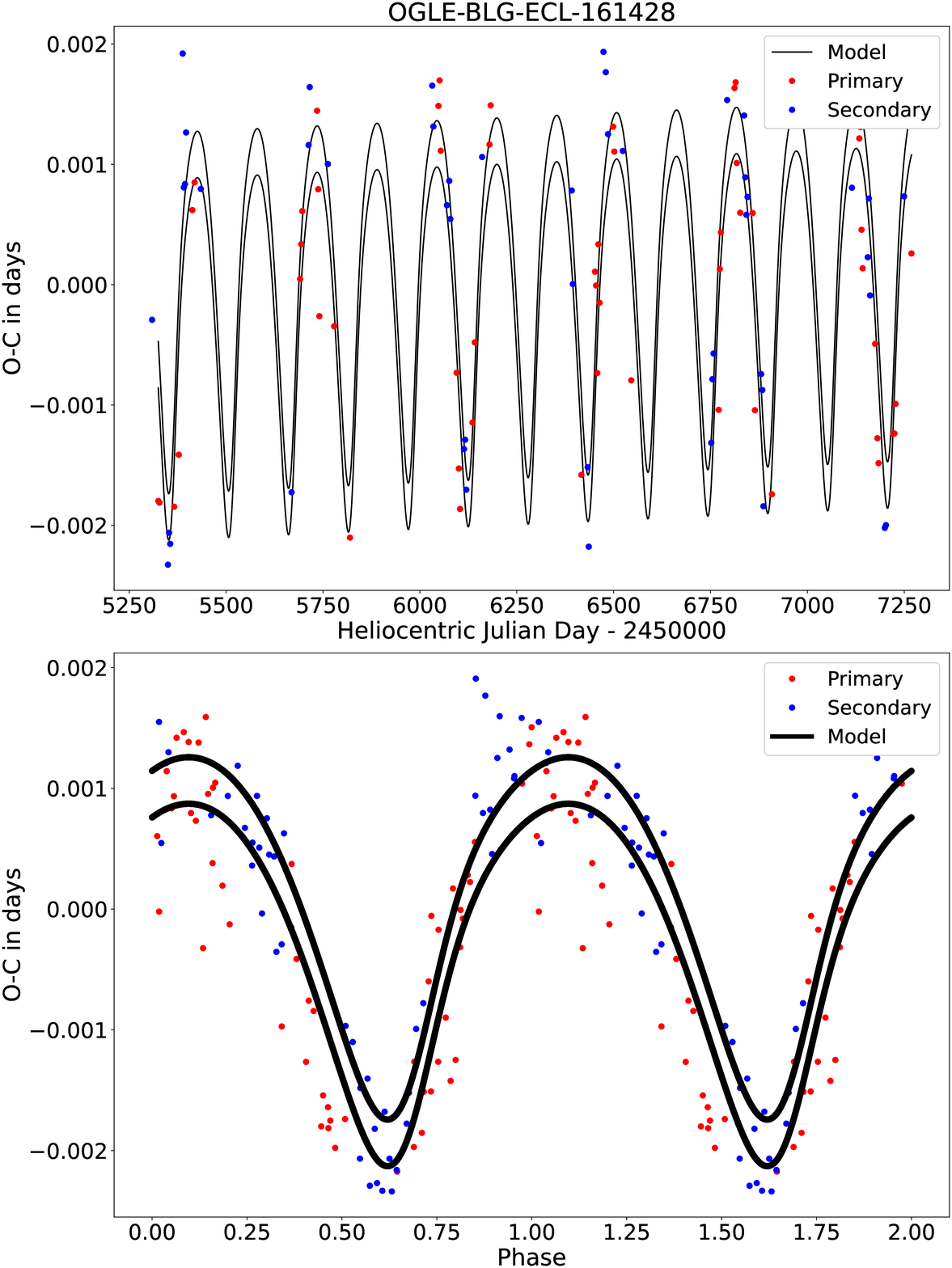}
    \includegraphics[width=\columnwidth]{Figs_new/199133_O-C_all.eps}
    
    \includegraphics[width=\columnwidth]{Figs_new/211268_O-C_all.eps}
    \includegraphics[width=\columnwidth]{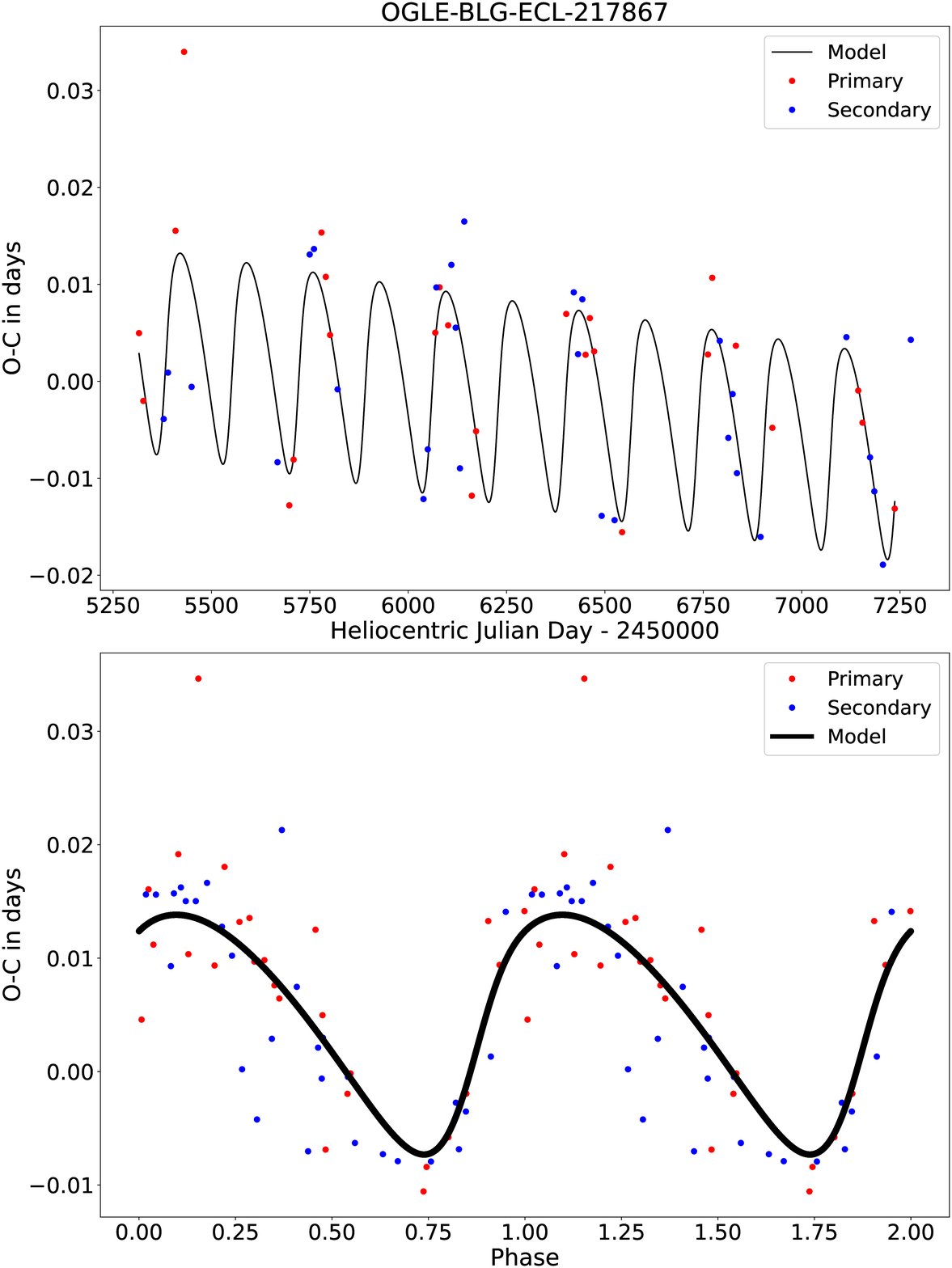}
    \clearpage
\end{figure*}

\begin{figure*}
    \centering
    \includegraphics[width=\columnwidth]{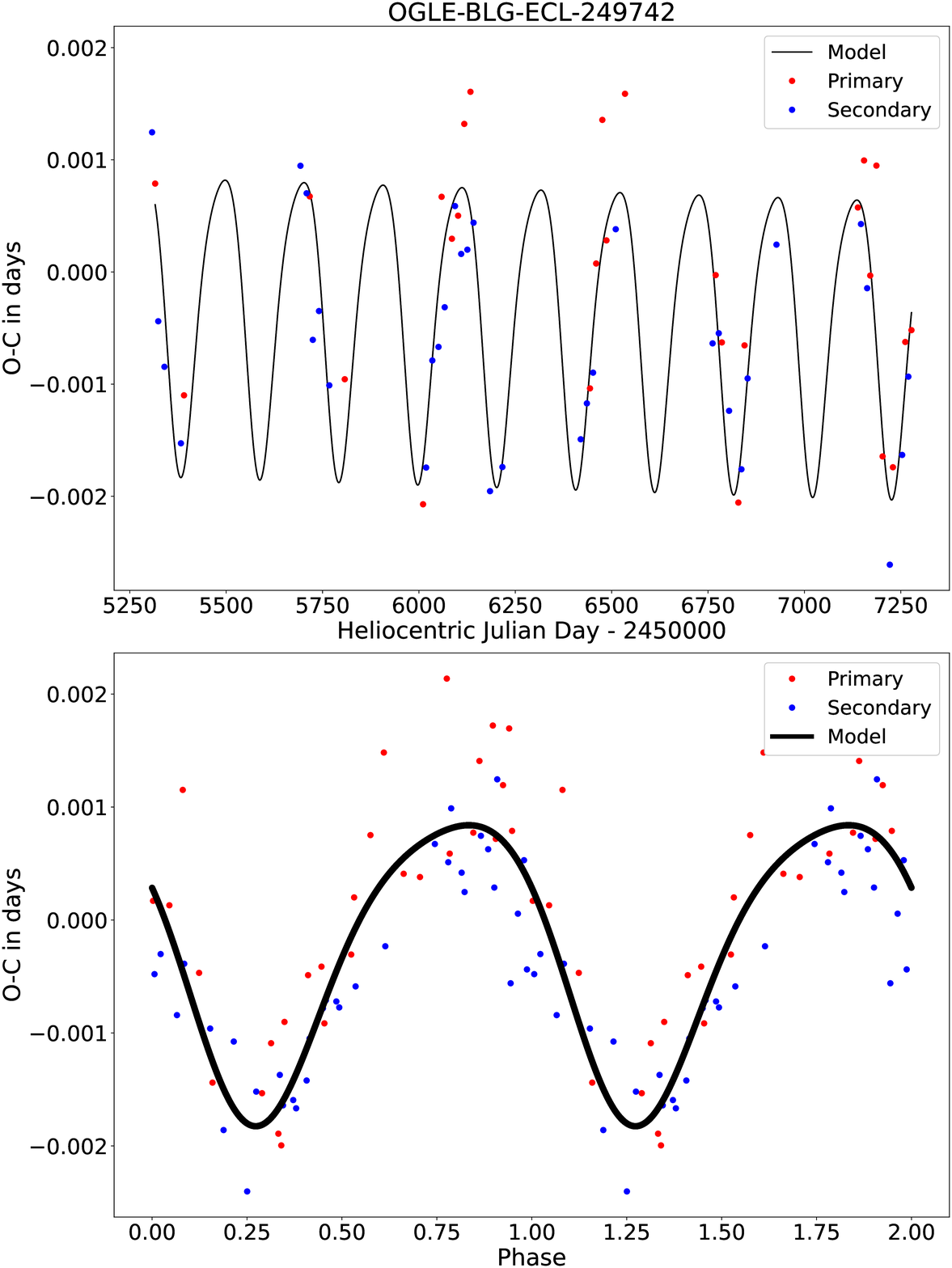}
    \includegraphics[width=\columnwidth]{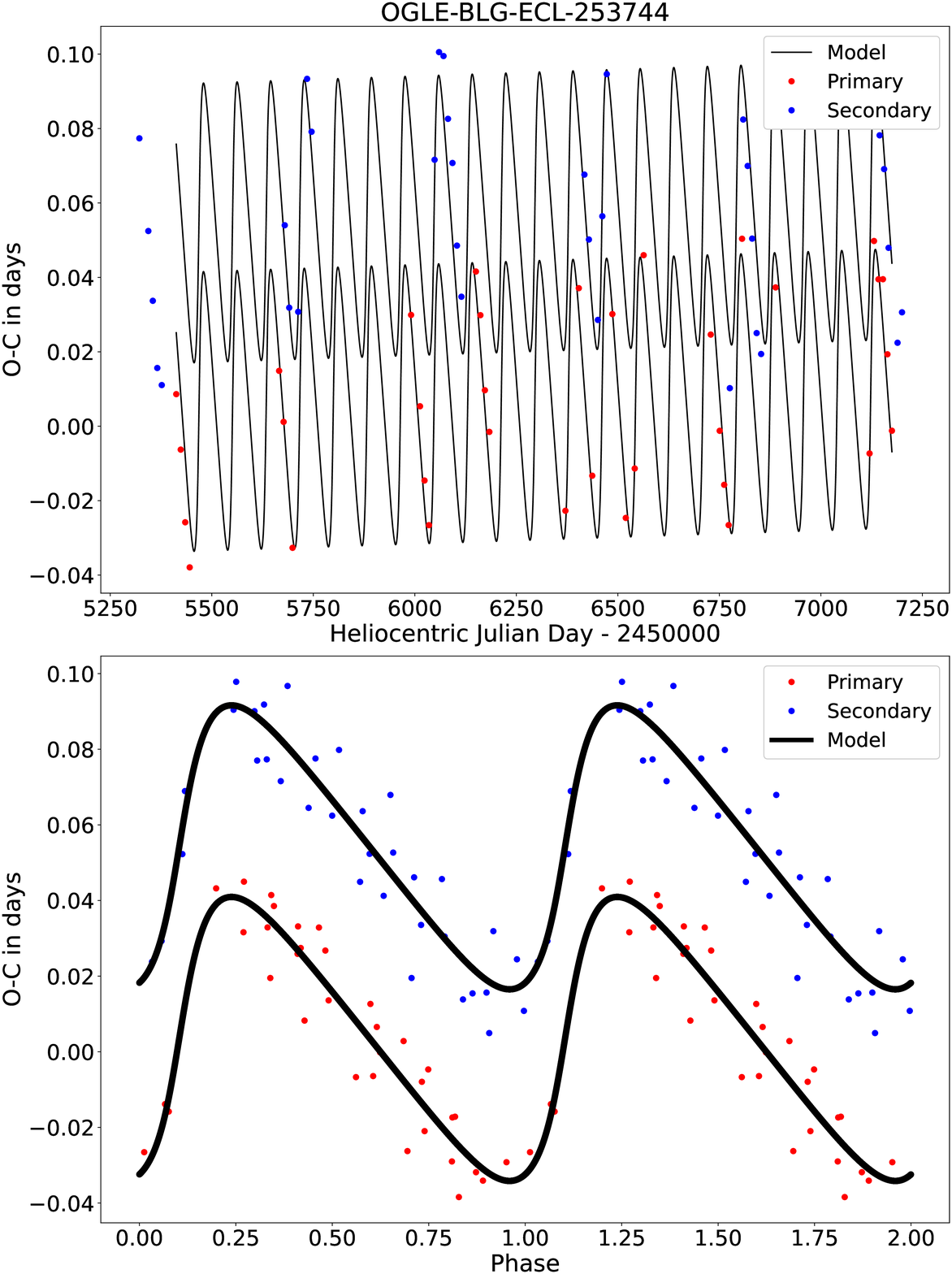}
\end{figure*}

\clearpage
\section{ETV of hierarchical triple stellar systems with LTTE solution}
\label{Appendix_B}

\begin{figure*}
    \centering
    \includegraphics[width=\columnwidth]{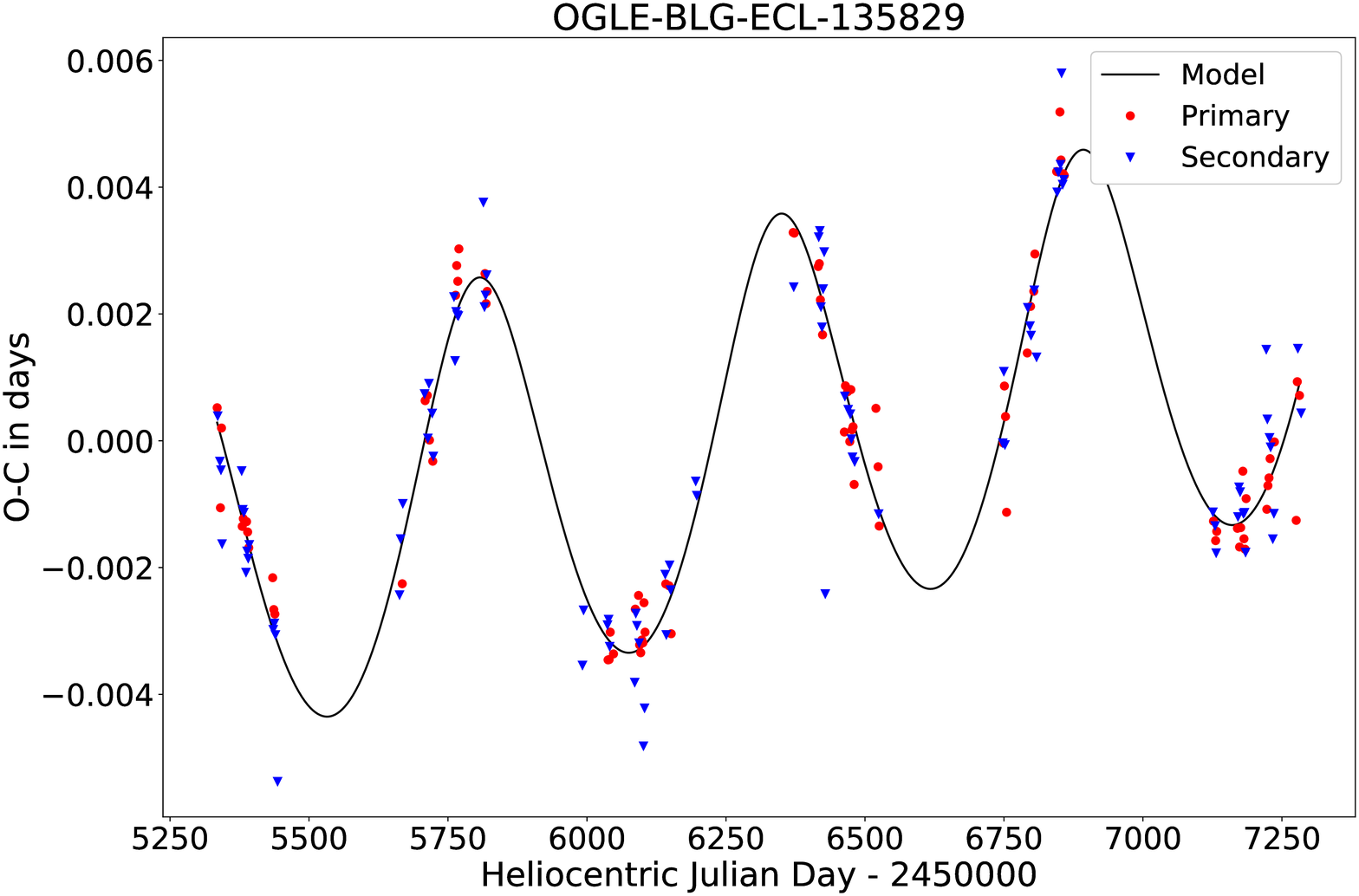}
    \includegraphics[width=\columnwidth]{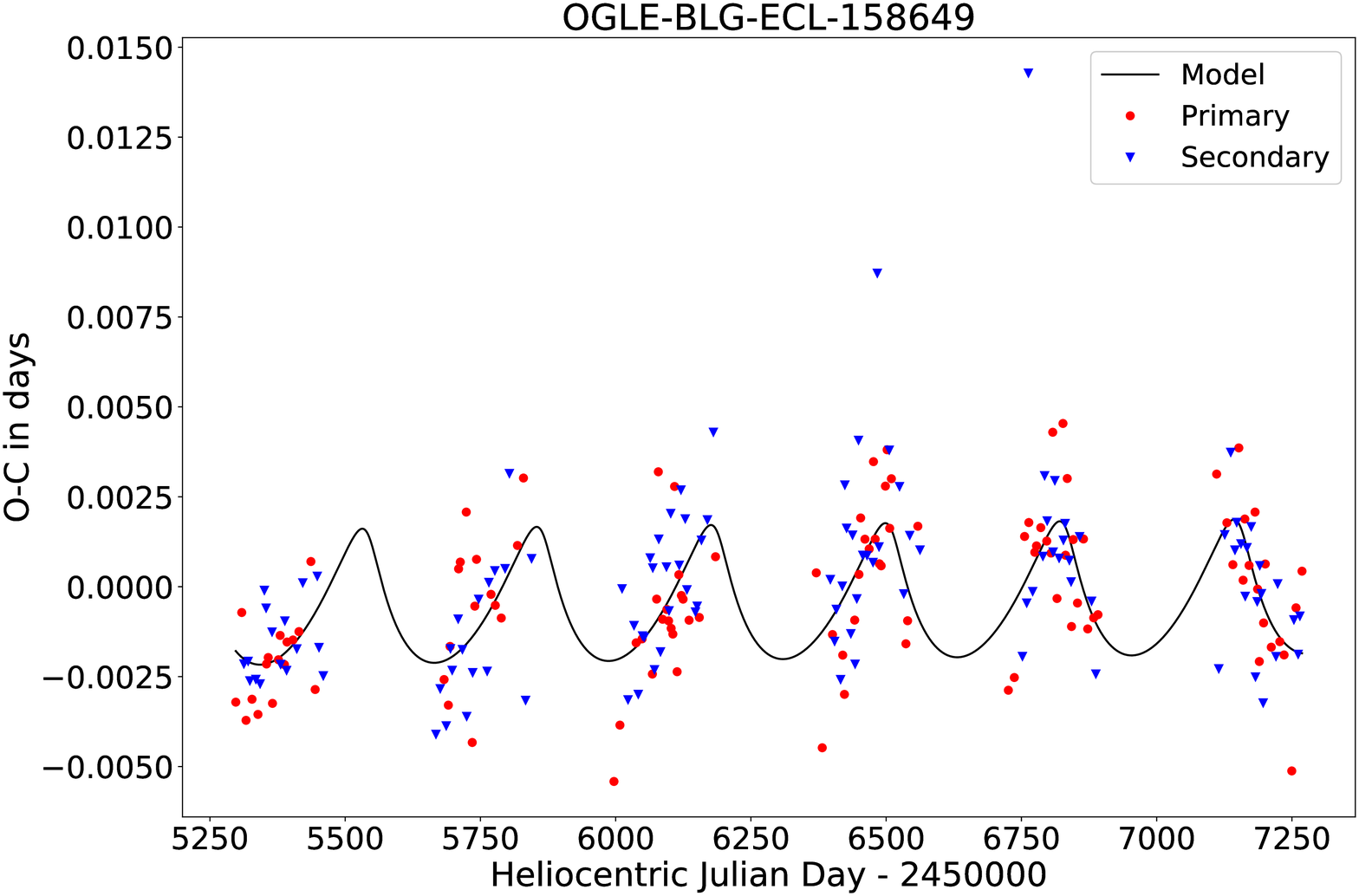}
    


    \includegraphics[width=\columnwidth]{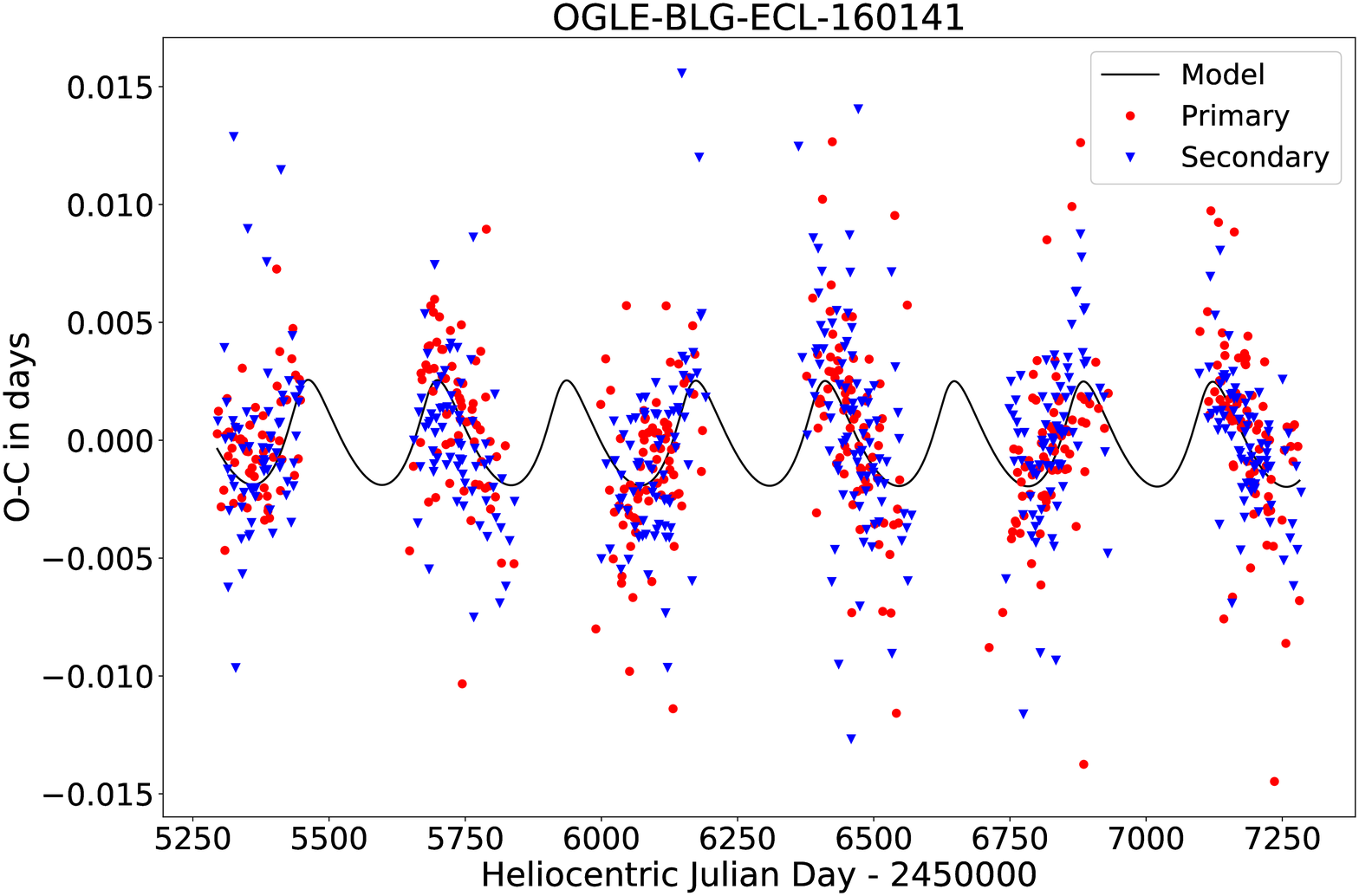}
    \includegraphics[width=\columnwidth]{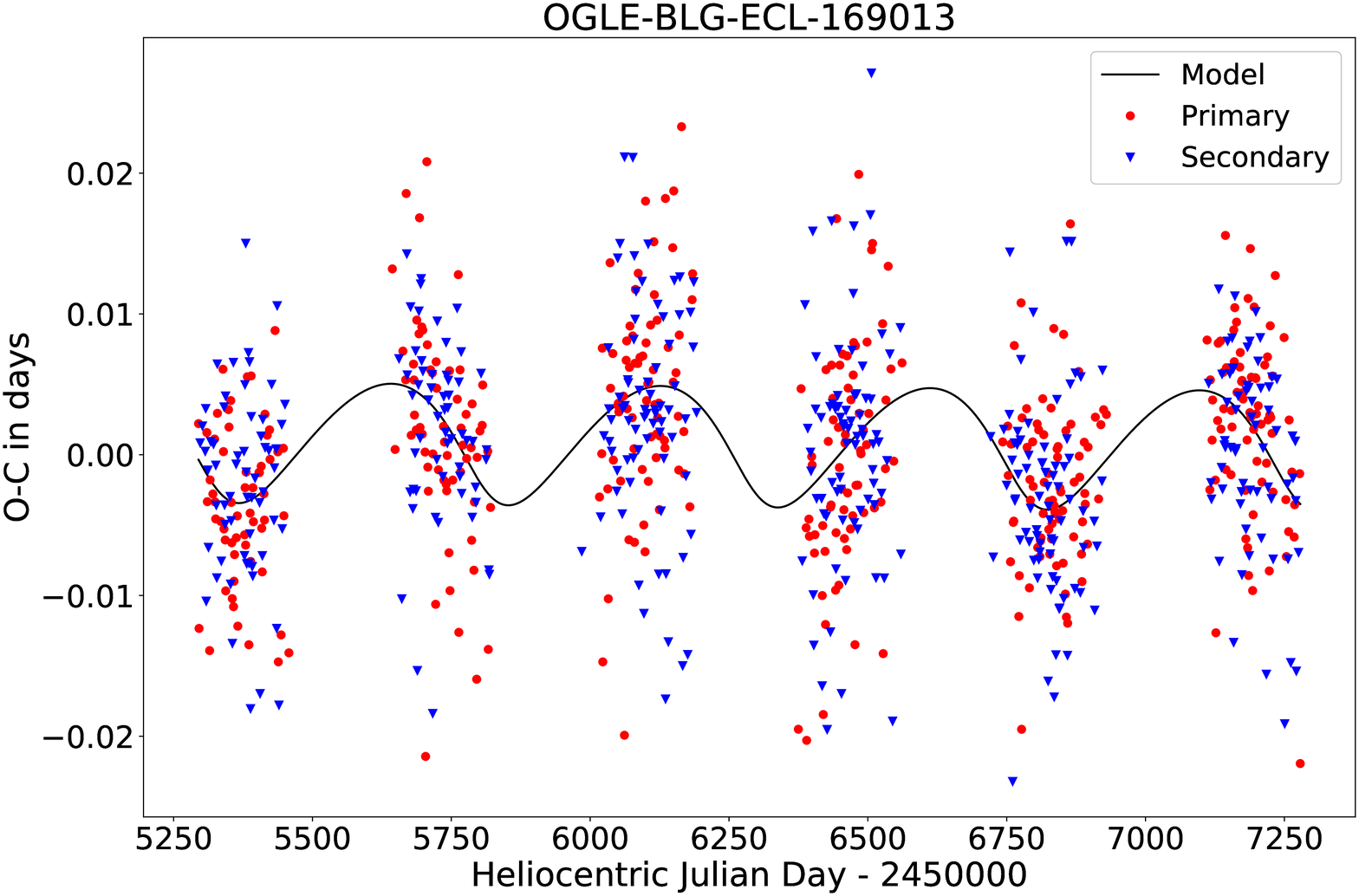}
    
    \includegraphics[width=\columnwidth]{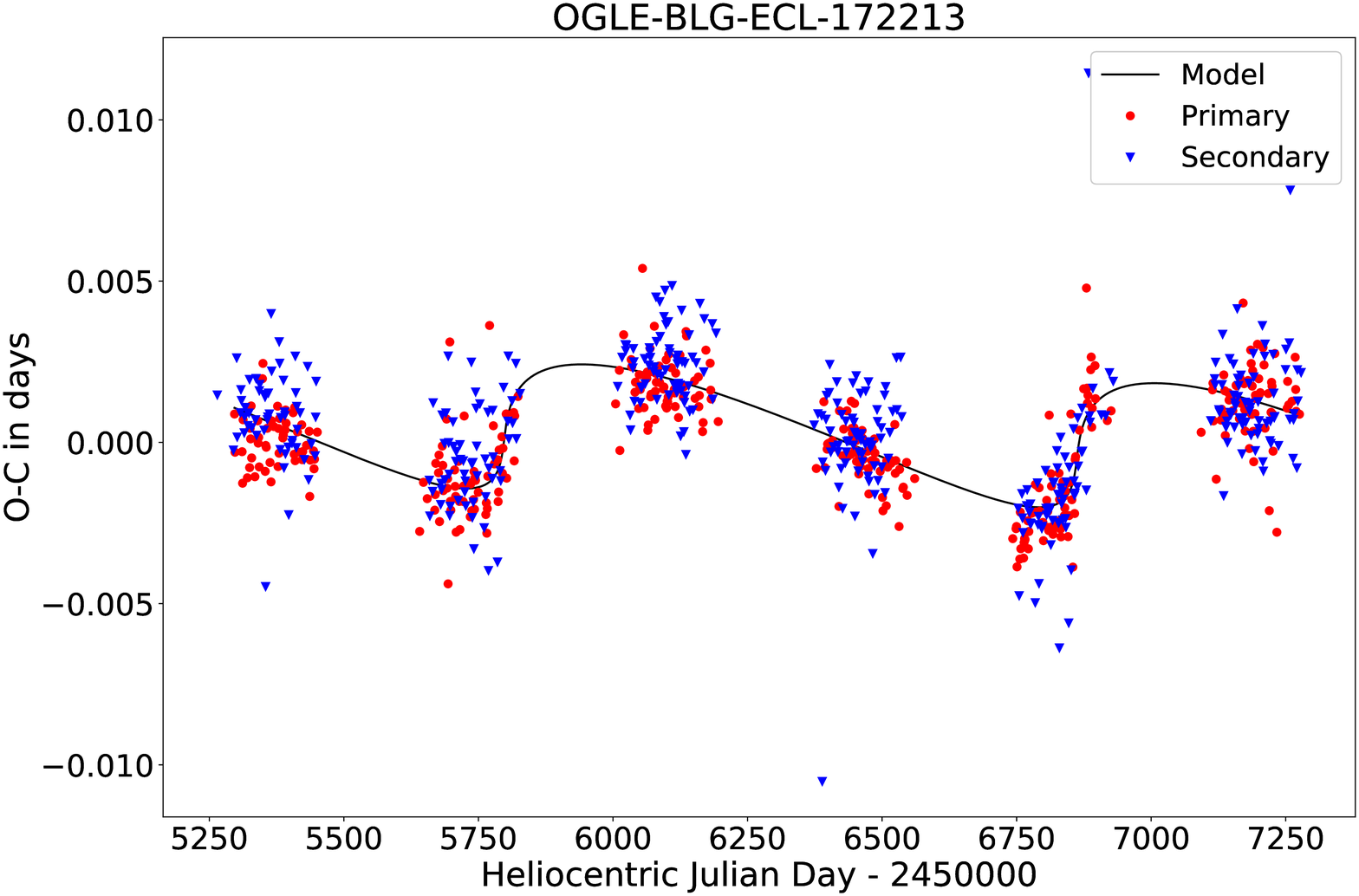}
    \includegraphics[width=\columnwidth]{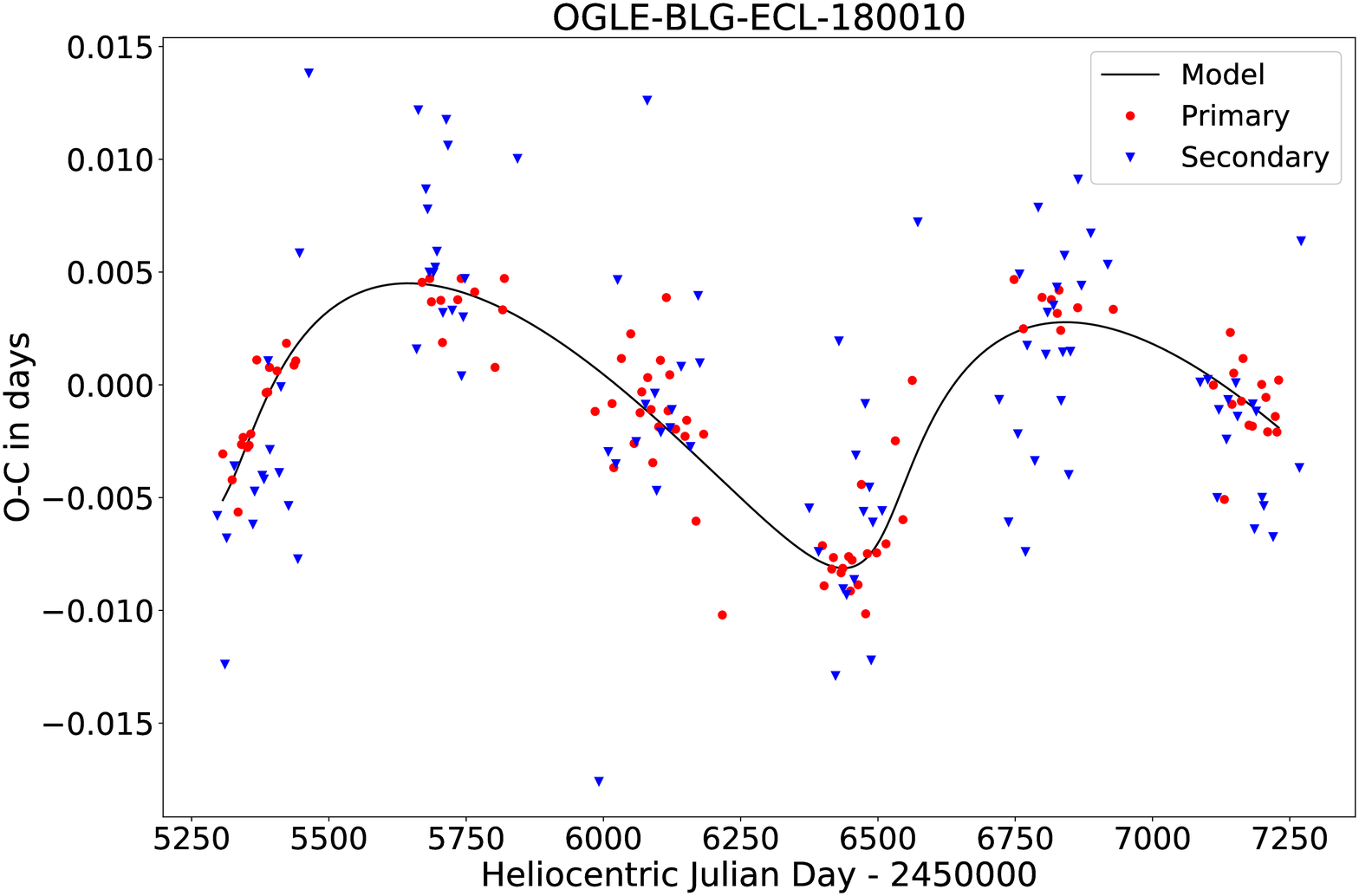}
\end{figure*}

\begin{figure*}
    \centering

    \includegraphics[width=\columnwidth]{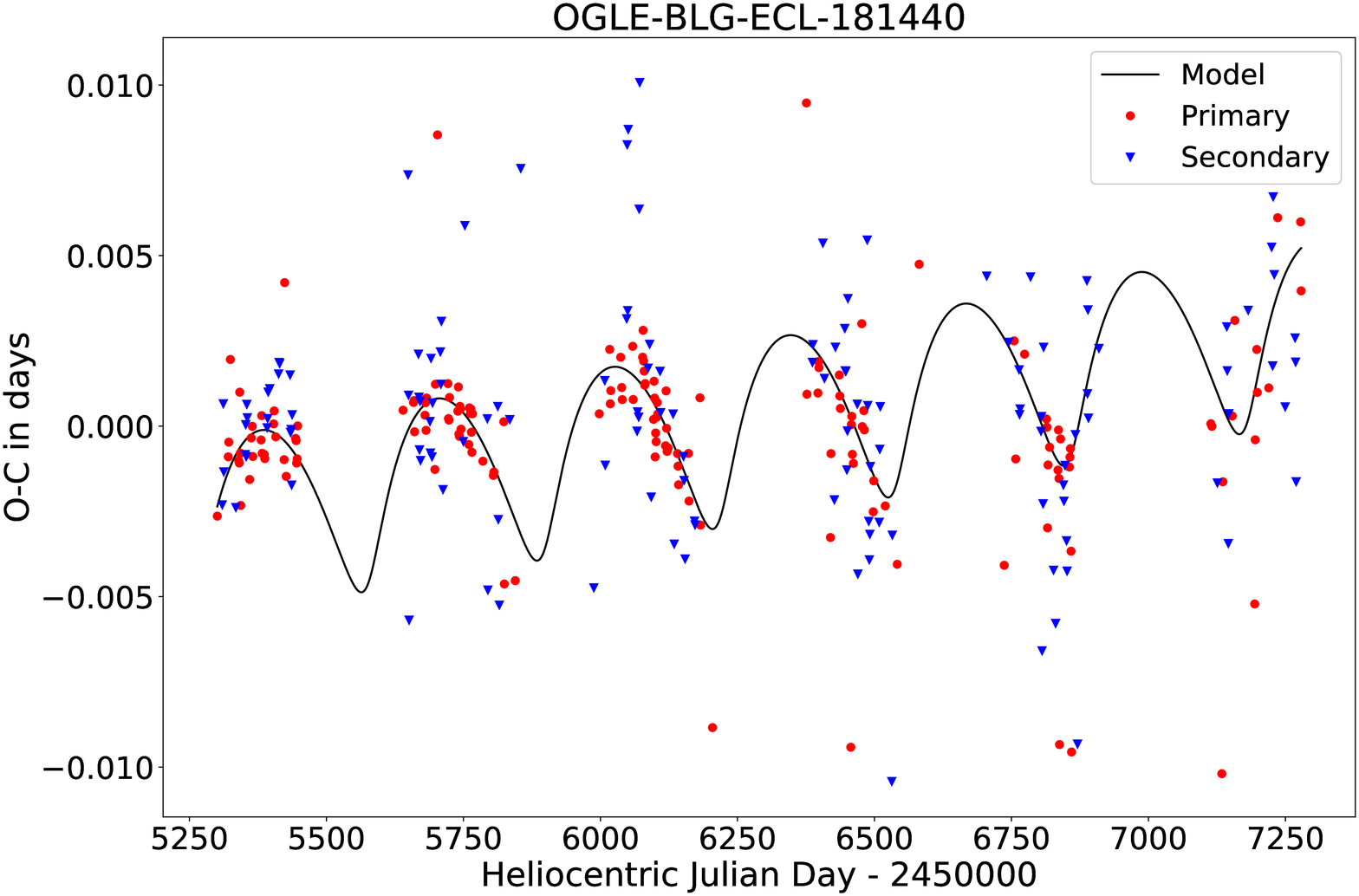}
    \includegraphics[width=\columnwidth]{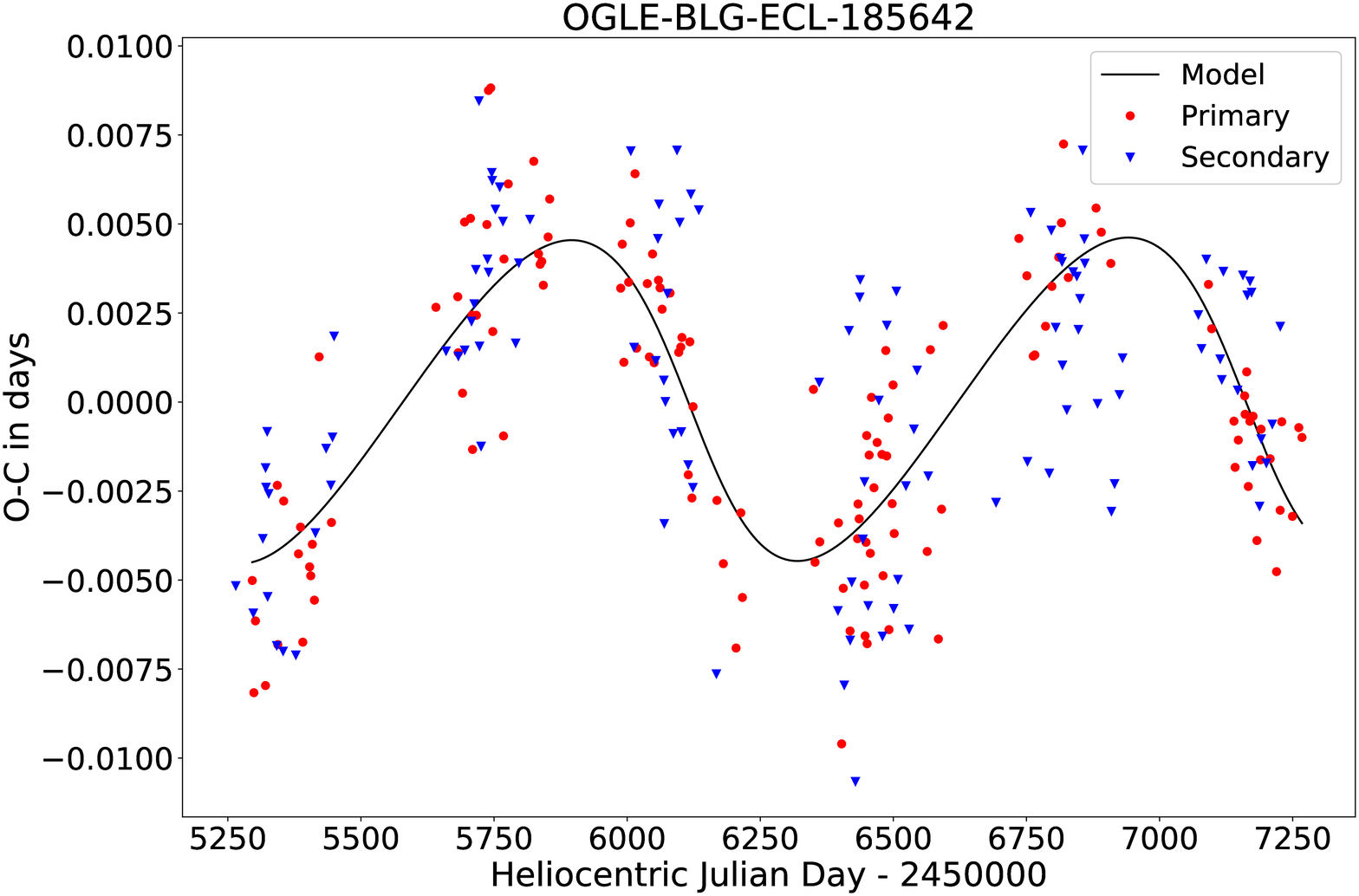}

    \includegraphics[width=\columnwidth]{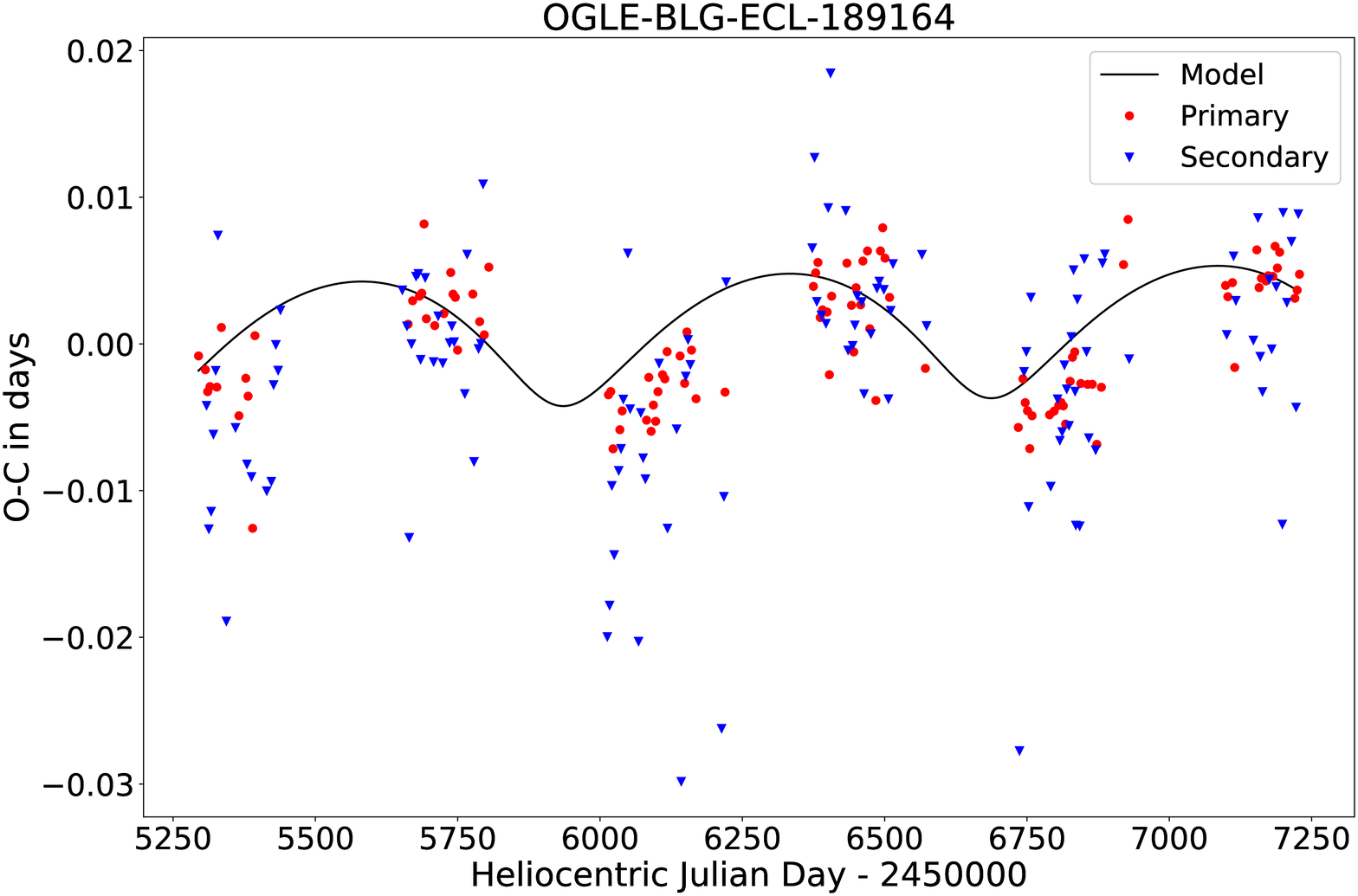}
    \includegraphics[width=\columnwidth]{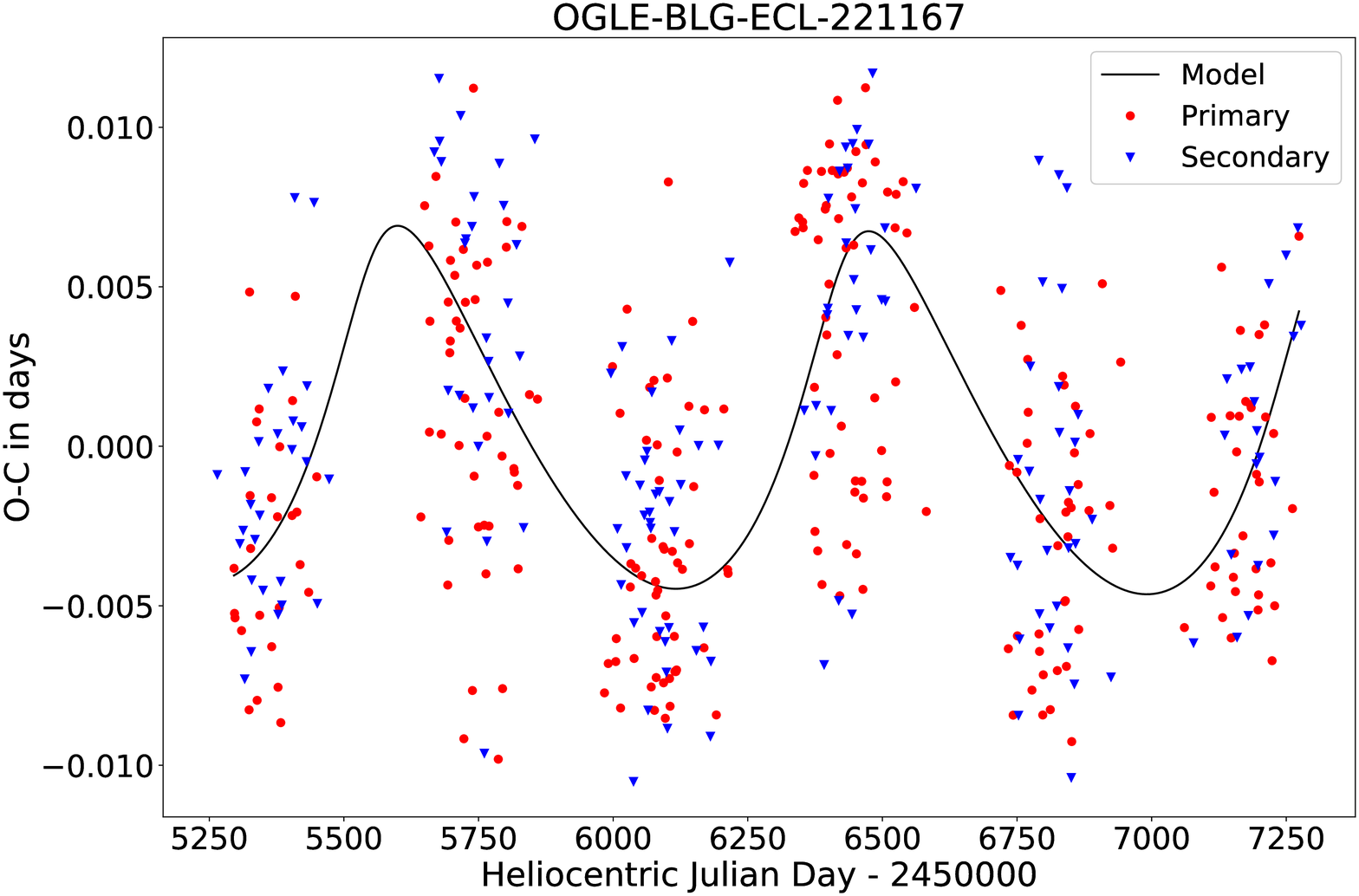}

    \includegraphics[width=\columnwidth]{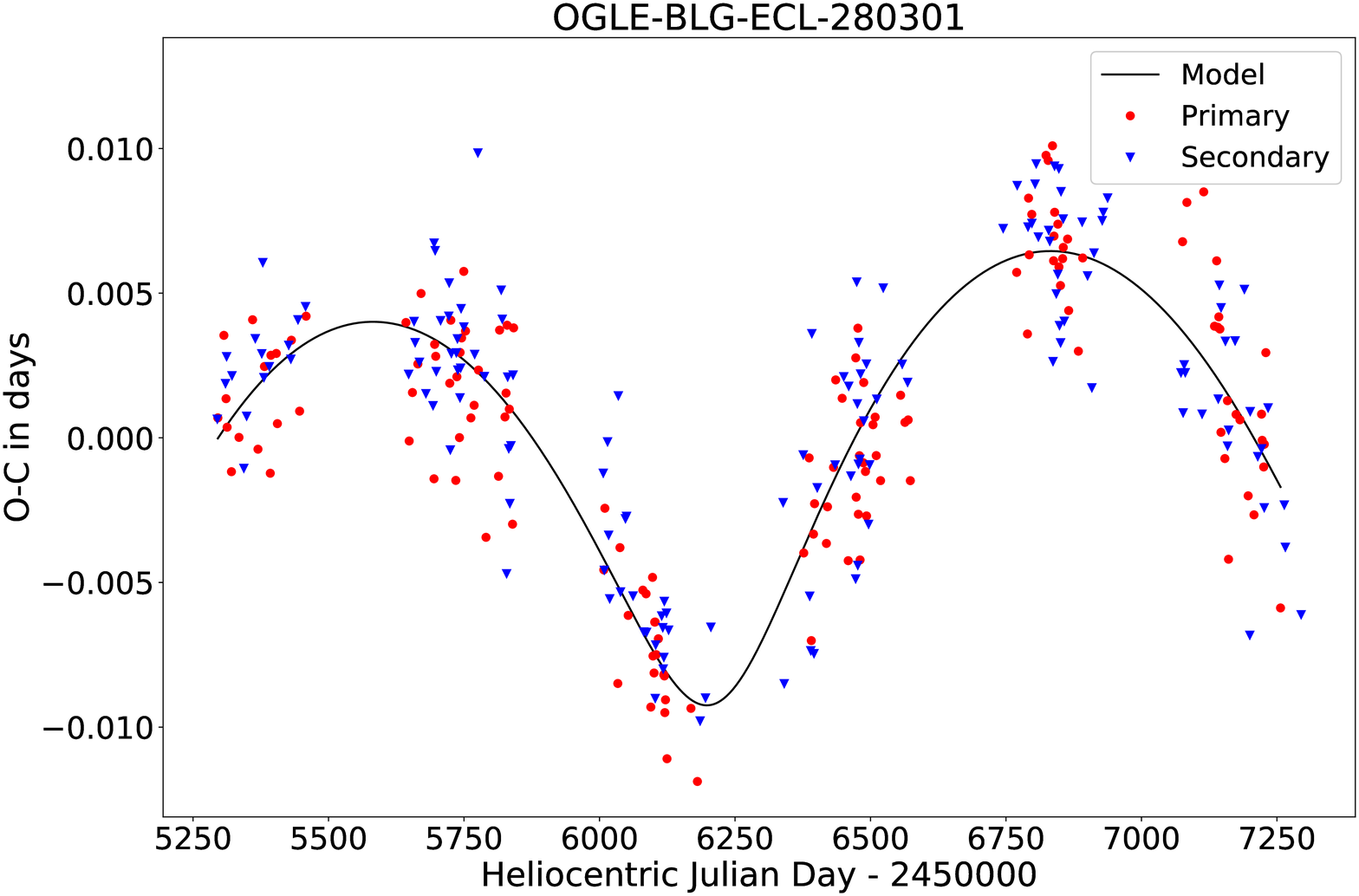}
    \includegraphics[width=\columnwidth]{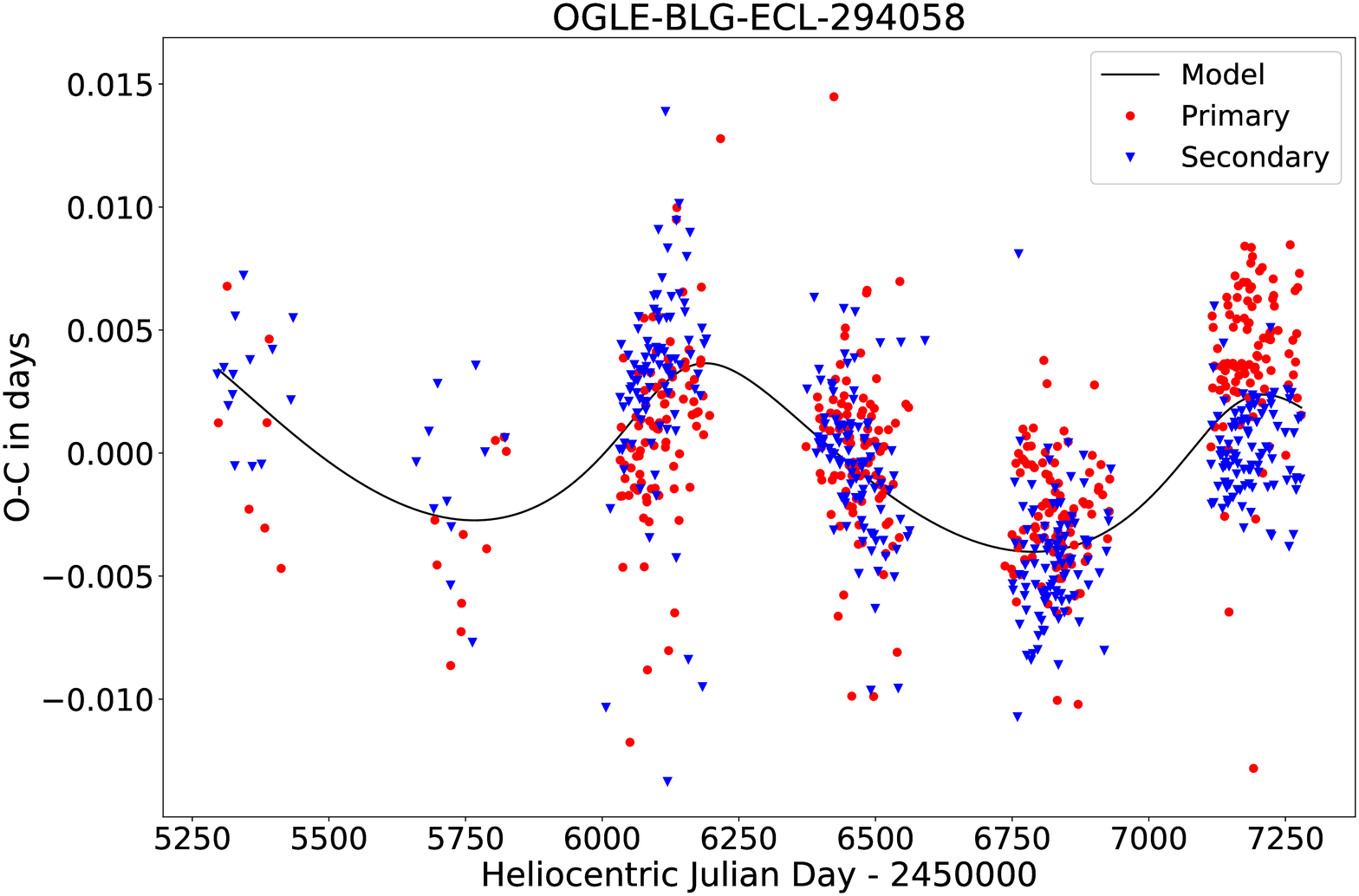}

\end{figure*}

\bsp	
\label{lastpage}
\end{document}